\documentclass[aps,pra,twocolumn,floatfix]{revtex4-1}

\usepackage{amsmath}
\usepackage{amssymb}
\usepackage{mathtools}
\usepackage{bbm}
\usepackage[caption=false]{subfig}
\usepackage[export]{adjustbox}
\usepackage{multirow}
\usepackage{natbib}

\usepackage{setspace}
\usepackage{bm}
\usepackage{graphicx}
\numberwithin{equation}{section}
\usepackage{hyperref}

\makeatletter
\renewcommand*\env@matrix[1][*\c@MaxMatrixCols c]{
  \hskip -\arraycolsep
  \let\@ifnextchar\new@ifnextchar
  \array{#1}}
\makeatother

\usepackage{amsmath, amsthm, amssymb}

\newcommand{\be}{\begin{equation}}
\newcommand{\ee}{\end{equation}}
\newcommand{\bea}{\begin{eqnarray}}
\newcommand{\eea}{\end{eqnarray}}

\usepackage{rotating,wrapfig}
\usepackage{footnote}
\usepackage{tcolorbox}
\newtcolorbox{mymathbox}[1][]{colback=white, sharp corners, #1}
\usepackage{empheq}
\usepackage{upgreek}

\graphicspath{ {threelevelsystem/},{intermediateplane/} ,{polaritoninteractions/}}

\begin{document}

\title{Theory of Interacting Cavity Rydberg Polaritons}
\author{Alexandros Georgakopoulos$^1$}
\author{Ariel Sommer$^2$}
\author{Jonathan Simon$^1$}
\affiliation{$^1$Department of Physics and James Franck Institute, University of Chicago, Chicago, IL}
\affiliation{$^2$Department of Physics, Lehigh University, Bethlehem, PA}

\date{\today}

\begin{abstract}
Photonic materials are an emerging platform to explore quantum matter~\cite{Carusotto2013,Sommer2015} and quantum dynamics ~\cite{Peyronel2012}. The development of Rydberg electromagnetically induced transparency ~\cite{weatherill2008electromagnetically,Petrosyan2011} provided a clear route to strong interactions between individual optical photons. In conjunction with carefully designed optical resonators, it is now possible to achieve extraordinary control of the properties of individual photons, introducing tunable gauge fields~\cite{Schine2016} whilst imbuing the photons with mass and embedding them on curved spatial manifolds~\cite{Sommer2016}. Building on work formalizing Rydberg-mediated interactions between propagating photons~\cite{Gorshkov2011,Gullans2016}, we develop a theory of interacting Rydberg polaritons in multimode optical resonators, where the strong interactions are married with tunable single-particle properties to build and probe exotic matter. In the presence of strong coupling between the resonator field and a Rydberg-dressed atomic ensemble, a quasiparticle called the ``cavity Rydberg polariton'' emerges. We investigate its properties, finding that it inherits both the fast dynamics of its photonic constituents and the strong interactions of its atomic constituents. We develop tools to properly renormalize the interactions when polaritons approach each other, and investigate the impact of atomic motion on the coherence of multi-mode polaritons, showing that most channels for atom-polariton cross-thermalization are strongly suppressed. Finally, we propose to harness the repeated diffraction and refocusing of the optical resonator to realize interactions which are local in momentum space. This work points the way to efficient modeling of polaritonic quantum materials in properly renormalized strongly interacting effective theories, thereby enabling experimental studies of photonic fractional quantum Hall fluids and crystals~\cite{ Sommer2015, Umucalilar2014,Grusdt2013}, plus photonic quantum information processors and repeaters~\cite{Jaksch2000,Saffman2010,han2010quantum}.
\end{abstract}

\maketitle

\section{Introduction} \label{sec:intro}

Current efforts to produce and explore the properties of synthetic quantum materials take numerous forms, from ultracold atoms~\cite{Stormer1999} to superconducting circuits~\cite{Houck2012,Anderson2016,Roushan2017} and electronic heterostructures~\cite{Tsui1982,Deng2010} and superlattices~\cite{Dean2010}. Cold atom techniques allow for precise control through lattice tuning~\cite{Greiner2002} and Feshbach resonances~\cite{Inouye1998,Vuletic1999}. Superconducting quantum circuits present an opportunity to create materials from strongly interacting microwave photons, as they exhibit excellent coherence~\cite{Schreier2008}, strong interactions ~\cite{Schuster2007}, and have recently been shown to be compatible with low disorder lattices~\cite{Underwood2012}, low loss lattice gauge fields ~\cite{Owens2018}, and interaction \& dissipation driven phase transitions~\cite{Fitzpatrick2017}.

In parallel, there is now growing interest in creating materials from optical photons. Non-interacting photons have been Bose-condensed in a resonator using a dye as a thermalization medium~\cite{Klaers2010}; photons have been made to interact weakly and subsequently Bose condense by coupling them to interacting excitons~\cite{Deng2010}. To explore strongly interacting photonic materials, it has previously been proposed to marry Rydberg electromagnetically induced transparency (EIT) tools developed to induce free-space photons to interact~\cite{Carroll2004,Gorshkov2011,Peyronel2012,Firstenberg2013} with multimode optical resonators~\cite{Sommer2015} to control the properties of individual photons~\cite{Sommer2016}, thereby introducing a real mass for 2D photons, and effective magnetic fields~\cite{Schine2016}, in conjunction with Rydberg mediated interactions. It was recently experimentally demonstrated that individual cavity photons do indeed hybridized with Rydberg excitations to form ``cavity Rydberg polaritons,'' quasiparticles ~\cite{Parigi2012,stanojevic2013dispersive,Jia2016} that collide with one another with high probability ~\cite{Jia2018}. 

Formal modeling of these complex systems is incomplete. The properties of interacting free-space Rydberg polaritons have been explored in the dispersive regime ~\cite{Gullans2016}, as well as the resonant regime for Van der Waals ~\cite{Gorshkov2007c,Bienias2014} and dipolar interactions ~\cite{Lukin2001}. Effective models of strongly interacting two-level cavity polaritons have been developed~\cite{litinskaya2016cavity}, along with blockade ``bubble'' approximations that qualitatively reflect the physics of three-level polaritons~\cite{Grankin2014}, but to date no effective theories of three-level cavity Rydberg polaritons exist which quantitatively reproduce the observed strong interactions, as a consequence of the intricate renormalization of the two-polariton wavefunction once the polaritons overlap in space.

In this paper, we first show that cavity Rydberg polaritons at large separations are described by the Hamiltonian $H_{pol} \sim \cos^2 \frac{\theta_d}{2} \text{ } H_{phot}+\sin^4 \frac{\theta_d}{2} \text{ } H_{int}$, where $H_{phot}$ is the Hamiltonian describing the bare cavity-photon dynamics, determined through the resonator geometry; and $H_{int}$ is the Hamiltonian describing the Rydberg-Rydberg interactions. The polaritons thus inherit properties from both photonic and atomic constituents, with the proportion of each contribution determined by the dark state rotation angle $\theta_d$ ~\cite{fleischhauer2005electromagnetically}, providing an interaction tuning knob akin to an atomic Feshbach resonance ~\cite{Inouye1998}. In the remainder of the paper we examine the limitations of this model, providing quantitative refinements to various aspects of it.

In section~\ref{atomresonatorcoupling} we begin with the Floquet Hamiltonian for non-interacting resonator photons~\cite{Sommer2016} and formally couple these photons to an ensemble of Rydberg-dressed three-level atoms residing in a waist of the resonator~\cite{Sommer2015}. In section~\ref{polaritonbasis} we explore the physics of an individual photon in the resonator, discovering one long lived dark polariton (with renormalized mass relative to the bare photon) and two short-lived bright polaritons. In section~\ref{polaritonpolaritoninteractions} we generalize to the case of two dark polaritons in the resonator, derive the form of the low-energy polariton-polariton interaction potential, investigate scattering into bright-polariton manifolds as well as the regime of validity of the two polariton picture in the face of interactions and loss, focusing in section~\ref{interactiondrivenloss} on collisional loss of polaritons by dark$\rightarrow$bright scattering. Once the interaction energy becomes larger than the dark/bright splitting the simple polaritonic picture breaks down, so in section~\ref{effectivetheory} we explore the maximally challenging case of two dark polaritons in single mode optical resonator, developing a properly renormalized effective theory of interacting polaritons (with first principles calculable parameters) that we benchmark against a complete (and numerically expensive) microscopic theory. We find excellent agreement in experimentally relevant parameter regimes, pointing the way to a fully renormalized effective field theory of multimode cavity Rydberg polaritons. In section~\ref{offplaneinteractions} we demonstrate that a properly situated Rydberg-dressed atomic ensemble produces interactions between polaritons that are local in momentum-space. In section~\ref{dopplerdecoherence} we relax the assumption of stationary atoms and investigate the effect of atomic motion on polariton coherence in both a single- and multi- mode regimes. Finally, in section~\ref{outlook} we conclude with a discussion of applications of cavity Rydberg polaritons to quantum information processing and strongly-correlated matter.

\begin{center}
 \begin{tabular}{|c|c|c|} 
 \hline
 \multicolumn{3}{|c|}{Table of Results} \\
 \hline 
  & Equation & Page \\
 Polariton Projected Hamiltonian & \ref{totalhamiltonian} & \pageref{totalhamiltonian}  \\ 
 \hline
 Renormalized Theory & \ref{collectiveRRstate},~\ref{effectiveinteraction},~\ref{effectiveOmega} & \pageref{collectiveRRstate} \\
 Role of Atomic Motion & \ref{timedependenthamiltonian} & \pageref{timedependenthamiltonian} \\
 \hline
 Doppler Decoherence & \ref{dopplershift1} & \pageref{dopplershift1} \\
 \hline
 Crossmode Thermalization & \ref{dopplerbroadening1} & \pageref{dopplerbroadening1} \\
 \hline
\end{tabular}
\end{center}

\section{Coupling the Photons to an Atomic Ensemble} \label{atomresonatorcoupling}
Here we explore how the coupling to an ensemble of three-level atoms impacts the physics of non-interacting 2D resonator photons. We find the emergence of long lived ``dark'' polaritons, with dynamics similar to those of a resonator photon but renormalized mass and harmonic trapping. Off-resonant, nonadiabatic couplings to ``bright'' polaritons limit the lifetime of the dark polaritons. We operate in the limit that the light-matter coupling energy scale is much larger than the energy scale of the photonic dynamics within the resonator, making the polaritonic quasiparticles a nearly ``good'' basis for describing the physics, with corrections that we derive.

The second quantized Hamiltonian for photons within a single longitudinal manifold of a resonator is given by~\cite{Sommer2016}:

\begin{eqnarray}
H_{phot}=\int \mathrm{d}x \textrm{ }a^{\dagger{}}(x)h_{phot}a(x), \nonumber 
\end{eqnarray}

\noindent where $a^{\dagger{}}(x)$ creates a photon at transverse location $x$ and $h_{phot}$ is the single particle Hamiltonian for a photon within the resonator, typically given by $h_{phot}(x)=\frac{\mathbf{\Pi}^2}{2m_{phot}}+\frac{1}{2}m_{phot}\omega_{trap}^2|x|^2-i\frac{\kappa}{2}$. Here $x$ runs over the plane transverse to the resonator axis, $\kappa$ parametrizes the (mode independent) resonator loss, and $\mathbf{\Pi}\equiv i\hbar\mathbf{\nabla}-e\mathbf{A}$ is the mechanical momentum; the parameters of this photonic ``Floquet'' Hamiltonian are determined by resonator geometry: mirror locations and curvatures, plus the twist of the resonator out of a single plane~\cite{Sommer2016,Schine2016}.

\begin{figure}[ht]
     \centering
     \subfloat[][]{\includegraphics[scale=0.325,valign=c]{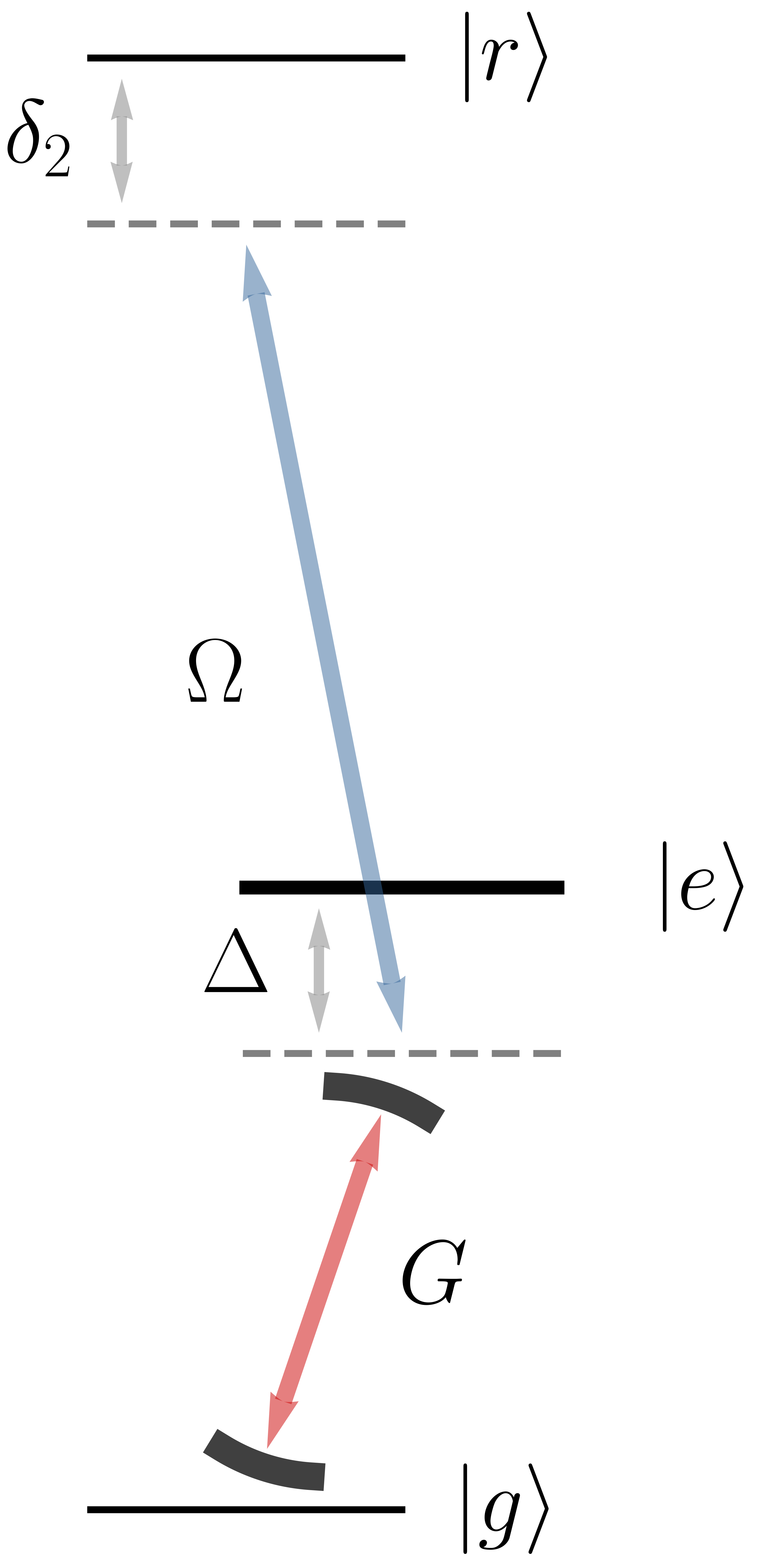}\label{fig:threelevelsystemfig}}
     \subfloat[][]{\includegraphics[scale=0.35,valign=c]{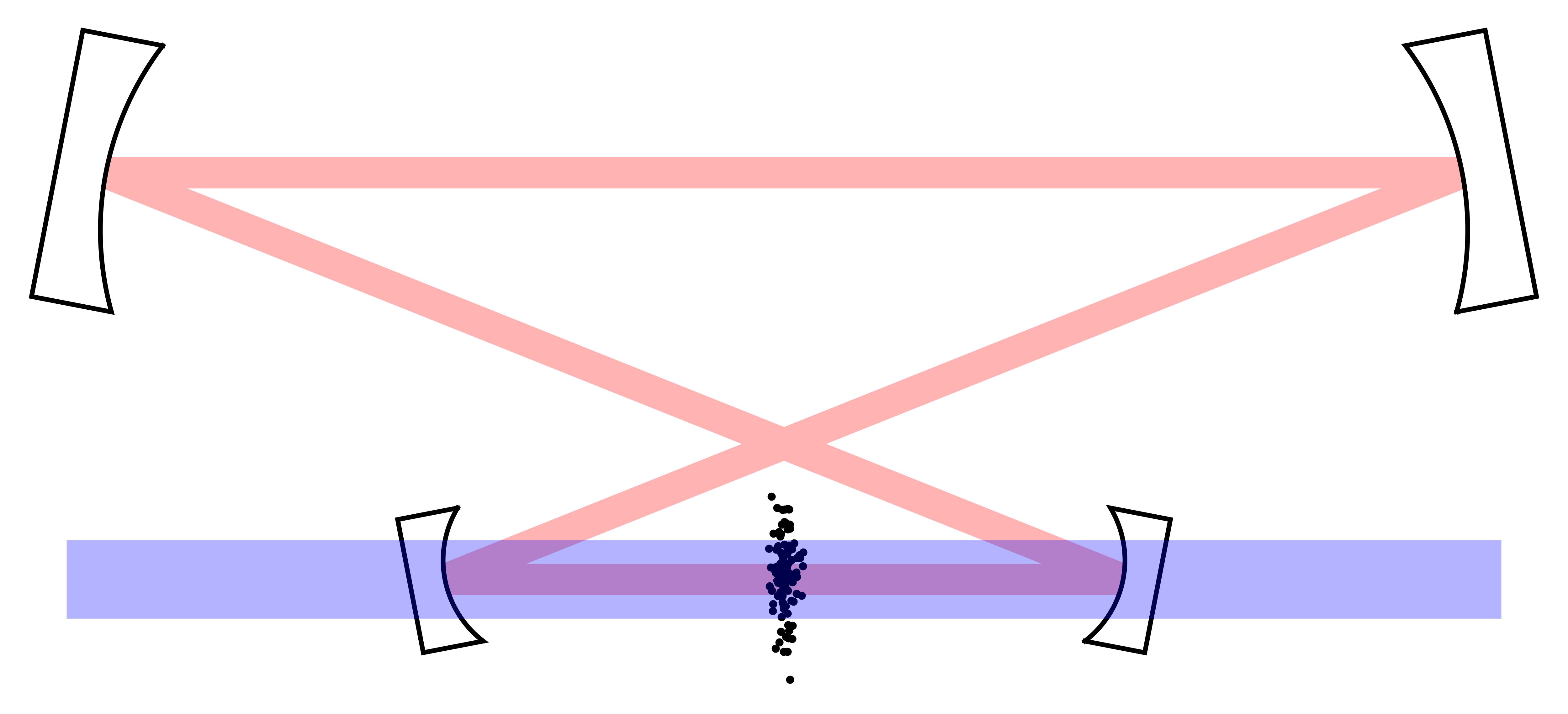}\label{fig:fourmirrorcavity}}
     \caption{\textbf{Resonator Rydberg polariton three-level system and setup schematic. (a)} The atomic ground state $|g\rangle$ is coupled to the excited state $|e\rangle$ through the cavity mode, with detuning $\delta_e$ and collective coupling strength $G$. The control beam then excites the atoms to the Rydberg state $|r\rangle$, with Rabi frequency $\Omega$ and two-photon detuning $\delta_2$.The horizontal displacement of the energy levels indicates a change in circular polarization. \textbf{(b)} The four mirror resonator supports running wave modes and two waists, allowing for both real and momentum space interactions. The primary (real space) atomic ensemble sits in the lower waist, where it mediates photon-photon interactions by admixing in a Rydberg state using the blue control field.}
     \label{steady_state}
\end{figure}

We now insert a Rydberg-dressed atomic ensemble into the resonator as a tool to mediate interactions between the photons. To this end, the lower (S$\rightarrow$P) transition of this ensemble is coupled to the quantized resonator field, while the upper (P$\rightarrow$Rydberg) transition is coupled to a strong coherent field (see Fig.~\ref{fig:threelevelsystemfig}). Before exploring the resulting photon-photon interactions, we must first understand how the Rydberg-dressed atoms impact the linear dynamics of individual photons. The light-matter coupling induced by the introduction of the atomic ensemble takes the form (in the frame rotating with the resonator- and Rydberg-dressing fields):

\begin{eqnarray}
H_{at}=\int \mathrm{d}x \textrm{ } \{ \phi^{\dagger{}}_r(x,z) \phi_r(x,z) \left (\delta_2-i\frac{\gamma_r}{2} \right )  \nonumber \\
+\phi^{\dagger{}}_e(x,z) \phi_e(x,z) \left (\delta_e-i\frac{\gamma_e}{2} \right ) \label{atomichamiltonian} \\
+\phi^{\dagger{}}_r(x,z) \phi_e(x,z)\frac{\Omega}{2}(x,z)+\mathrm{h.c.} \nonumber \\
\phi^{\dagger{}}_e(x,z)a(x)\frac{G(x,z)}{2}  +\mathrm{h.c.} \}.  \nonumber
\end{eqnarray}

Here $\phi^{\dagger{}}_r(x,z)$ and $\phi^{\dagger{}}_e(x,z)$ are the bosonic creation operators for atomic excitations at 3D location (x,z) in the Rydberg and excited (P) states respectively, from the ``vacuum'' of ground state atoms. $\gamma_r$ and $\gamma_e$ are the FWHM of the Rydberg and excited states respectively. $\delta_e$ is the detuning of an untrapped, zero-transverse-momentum resonator photon from the atomic line and $\delta_2$ is the two-photon detuning from EIT resonance. $\Omega(x,z)$ is the laser induced Rabi coupling between excited (P) and Rydberg states, while $G(x,z)$ is the vacuum-Rabi coupling strength between a resonator photon localized at transverse location $x$ and a collective atomic excitation localized at longitudinal location $z$, and therefore must reflect the atom density. It does not reflect the transverse spatial structure of any particular resonator mode, as $a^{\dagger{}}(x)$ creates a transversely localized photon. Indeed it may be written as $G(x,z)\approx d_{ge}\sqrt{\frac{1}{L_{res}}\frac{\rho(x)\hbar\omega_{ge}}{\epsilon_0}}$, where $L$ is the length of the atomic ensemble along the resonator axis, $L_{res}$ is the length of the resonator itself, $d_{ge}$ is the dipole moment of the atomic transition coupled to the optical resonator, $\omega_{ge}$ is the angular frequency of this transition, and $\rho(x)$ is the number density of atoms at location $x$.

Here and throughout, we incorporate losses through non-Hermitian Hamiltonians rather than through Lindbladian master equations. This allows us to identify the imaginary part of one- and two- particle eigenstates with particle decay rates ~(\cite{cohen1998atom} III.B.2).

\section{Polariton Basis} \label{polaritonbasis}

The atomic density distribution exists in three dimensions, while the manifold of nearly-degenerate resonator modes to which the atoms couple is two-dimensional (here we assume $G\ll\frac{c}{L_{res}}$, ensuring that the atoms couple to only a single longitudinal manifold of the resonator). In order to develop a formalism of two-dimensional polaritons, we define longitudinally delocalized, transversely localized collective atomic excitation operators (for $k_{laser}$ and $k_{cav}$ the wave-vector magnitudes of the coupling-laser and resonator fields, respectively), normalized to ensure a mode-independent bosonic commutation relation. To this end we choose the minimal case of an atomic ensemble of uniform density, and a uniform coupling-field that propagates counter to the resonator field:

\begin{align}
\phi^{\dagger{}}_e(x) \equiv \sqrt{\frac{1}{L}} \int_{-\frac{L}{2}}^{\frac{L}{2}} \mathrm{d}z \text{ } \phi^{\dagger{}}_e(x,z)e^{-i k_{cav} z} \nonumber \\
\phi^{\dagger{}}_r(x) \equiv \sqrt{\frac{1}{L}} \int_{-\frac{L}{2}}^{\frac{L}{2}} \mathrm{d}z \text{ } \phi^{\dagger{}}_r(x,z)e^{i(k_{laser}-k_{cav}) z}. \nonumber
\end{align}

We can now rewrite the atomic Hamiltonian as:

\begin{eqnarray}
H_{at}=\int \mathrm{d}x \textrm{ } [W^{\dagger{}}(x)][h_{at}][W(x)] \nonumber \\
\text{where } [W^{\dagger{}}] \equiv (a^{\dagger{}}(x) \text{ } \phi_e^{\dagger{}}(x)  \text{ } \phi_r^{\dagger{}}(x)) \nonumber \\
\text{and } [h_{at}]=\begin{pmatrix}
0 & \frac{G}{2} & 0\\
\frac{G}{2} & \delta_e-i\frac{\gamma_e}{2} & \frac{\Omega}{2}\\
0 & \frac{\Omega}{2} & \delta_2-i\frac{\gamma_r}{2}
\end{pmatrix}. \nonumber
\end{eqnarray}

\noindent Here $G\approx d_{ge}\sqrt{\frac{L}{L_{res}}\frac{\rho\hbar\omega_{ge}}{\epsilon_0}}$, where $L$ is the length of the atomic ensemble along the resonator axis. Note that $[h_{at}]$ has no position dependence. In the basis where $[h_{at}]$ is diagonal, the resulting creation operators are the generators of three varieties of polaritons: one dark (with little to no excited state participation, depending on $\kappa$ and $\gamma_r$), sandwiched between two bright (with large excited state participation). We are primarily concerned with the long-lived and strongly interacting dark polaritons, but will include off-resonant couplings to the bright polaritons to accurately model dark-polariton lifetime.

We diagonalize $[h_{at}]=\sum \mu_m \tilde{\mu}_m\epsilon_m$, where $m$ is an element of [d,b$_-$,b$_+$], meaning [dark, lower bright, upper bright], $\epsilon_m$ is the energy of (ket) $\mu_m $ and (bra) $\tilde{\mu}_m$ and the $\mu_m,\tilde{\mu}_l$ satisfy the generalized orthogonality condition (due to the non-hermicity of $h_{at}$): $\tilde{\mu_l} \centerdot \mu_m=\delta_{lm}$. Note that because of the non-hermicity of $h_{at}$, $\tilde{\mu}_m \neq \mu_m^{\dagger{}}$.

The resulting polariton creation and annihilation operators are thus:

\begin{eqnarray}
\chi_j^{\dagger}(x)&=&[W^{\dagger}] \cdot \mu_j, \nonumber \\
\chi_j(x)&=&\tilde{\mu_j} \cdot [W]. \nonumber 
\end{eqnarray}

\noindent For notational convenience \emph{only} have we named the polariton creation/annihilation operators ``$\chi_j^{\dagger}(x)$'' and ``$\chi_j(x)$''; these operators are \emph{not} precisely Hermitian conjugates of one another, but are instead defined to preserve the bosonic commutation relations: $[\chi_j(x),\chi^{\dagger}_k(x')]=\delta_{jk}\delta(x-x')$. We may now write $H_{at}$ as:

\begin{eqnarray}
H_{at}=\sum_m \int \textrm{ } \mathrm{d}x \textrm{ } \chi_m^{\dagger}(x) \chi_m(x) \epsilon_m. \nonumber 
\end{eqnarray}

\noindent  The last step in writing $H_{tot}$ in the polariton basis is to decompose $a^{\dagger}(x), a(x)$ into polariton field operators:
\begin{eqnarray}
a^{\dagger}(x)&=&\sum_j c_j \chi_j^{\dagger}(x), \nonumber \\
 a(x)&=&\sum_j \tilde{c}_j \chi_j(x), \nonumber 
\end{eqnarray}

\noindent where the $c_j,\tilde{c}_j$ are elements of the inverse $\mu,\tilde{\mu}$ matrices: $c_j \equiv \mu^{-1}_{cav,j},\tilde{c}_j \equiv \tilde{\mu}^{-1}_{cav,j}$ and the index ``cav'' denotes the photonic slot of $\mu^{-1},\tilde{\mu}^{-1}$. $H_{tot}$ can now be written as:

\begin{eqnarray}
H_{tot}=\sum_{ij}\int \mathrm{d}x \textrm{ }  c_i \tilde{c}_j \chi^{\dagger}_i(x)h_{tot}^{ij}(x)\chi_j(x), \nonumber
\end{eqnarray}

\noindent where $h^{ij}_{tot}(x) \equiv \delta_{ij}\epsilon_i+h_{phot}(x)$.

We will operate in the limit that the difference of the eigenvalues of $h_{at}$ (the ``dark''-``bright'' spitting) is much larger than the spectrum of $h_{phot}$, making the $[d,b_+,b_-]$ basis that diagonalizes the atomic Hamiltonian a near-diagonal basis for the multimode system. This is equivalent to the statement that a particle oscillating in the trap is more accurately described as a polariton rather than a photon if the light-matter coupling is much greater than the transverse optical mode spacing. To first order in $h_{phot}/ \epsilon$, the Hamiltonian projected into the dark polariton manifold is:

\begin{eqnarray}
H^{pol}_{tot}=\int \mathrm{d}x \textrm{ } 
 c_{d} \tilde{c}_{d}\chi^{\dagger}_{d}(x)(\epsilon_{d}+h_{phot}(x))\chi_{d}(x). \nonumber
\end{eqnarray}

\noindent For $\gamma_r=\delta_2=0$: $\epsilon_d=0$ and $c_{d} \tilde{c}_{d}=\frac{\Omega^2}{\Omega^2+G^2}=\cos^2\frac{\theta_{d}}{2}$, where $\theta_{d}$ is the dark state rotation angle~\cite{fleischhauer2005electromagnetically}. We then have:

\begin{eqnarray}
H^{pol}_{tot}= \cos^2\frac{\theta_{d}}{2}\int \mathrm{d}x \textrm{ } \chi^{\dagger}_{d}(x) \nonumber \\
\left[ \frac{\mathbf{\Pi}^2}{2m_{phot}} + \frac{1}{2} m_{phot} \omega^2_{trap}|x|^2-\frac{i \kappa}{2}\right] \nonumber \\
\chi_{d}(x).  \label{noninteractingpolaritonhamiltonian}
\end{eqnarray}

\noindent Thus, we see that to lowest order in ratio of the photonic dynamics energy scale to the atomic coupling energy scale $\frac{\omega^2_{trap}}{\Omega^2+G^2}$, the atoms simply slow down all photonic dynamics and loss by a factor of $\cos^2\frac{\theta_{d}}{2}$. The dominant correction to this story is a second-order resonator ($h_{phot}$) -induced dark$\rightarrow$bright coupling, producing an effective Hamiltonian~\cite{cohen1998atom} (in the above limits):

\begin{eqnarray}
\delta H_{tot}\approx\sum_{j\in[b_-,b_+]} \int\mathrm{d}x \frac{H_{phot} \chi_j^\dagger(x) \chi_j(x) H_{phot}}{\epsilon_j-\epsilon_d}\nonumber
\end{eqnarray}

\noindent For a dark-polariton in an eigenstate with energy \\
$E=\frac{\Omega^2}{G^2+\Omega^2} \delta \ll \sqrt{\Omega^2+g^2}$,where $\delta$ is the detuning of the corresponding bare photon eigenstate from EIT resonance, the correction is largely imaginary (assuming for simplicity that $\delta_e=0,g,\Omega \gg \gamma_e$):

\begin{eqnarray}
\delta \Gamma_{pol} \equiv -i \langle \delta H_{tot} \rangle &\approx& 2\frac{\Omega^2}{\Omega^2+G^2} \frac{G^2}{\Omega^2+G^2} \frac{\delta^2}{\Omega^2+G^2}\gamma_e \nonumber \\
&=&2 \tan^2 \frac{\theta_d}{2} \frac{E^2}{\Omega^2+G^2} \gamma_e \label{eqn:ploss}
\end{eqnarray}

\noindent Thus we see that loss is maximized (at fixed $G^2+\Omega^2$) for a polariton which is equal parts photon and  Rydberg excitation, yielding $\delta \Gamma_{pol} \approx 2 \frac{E^2}{\Omega^2+G^2}\gamma_e$, a worst-case approximation we will employ for simplicity going forward.

Additional contributions to polariton loss arise from inhomogeneous broadening of the Rydberg manifold, e.g. atomic motion and electric field gradients (to which Rydberg atoms are susceptible due to their large DC polarizability ~\cite{Saffman2010}). Such processes generate a ladder of couplings from the collective Rydberg state with the symmetry of the resonator mode into modes orthogonal to it (which therefore bright). We explore how (random) atomic motion induces polariton decoherence in Sec ~\ref{dopplerdecoherence}; for inhomogeneous E-fields, the coupling rate to the bright manifold is $\gamma_b \approx \alpha E \delta E$, where $\alpha$ is the DC polarizability of the Rydberg state, $E$ is the DC electric field at the atomic sample, and $\delta E$ is the field-variation across it. The resultant broadening of the dark manifold is then (in the limit $\Omega \gg \gamma_e$) ~\cite{Jia2016}:

\begin{eqnarray}
\delta \Gamma_{pol} \sim \gamma_e \frac{\gamma_b^2}{\Omega^2}.
\end{eqnarray}

It is instructive to compare this result with the loss induced by detuning a resonator mode out of the EIT window (Eqn.~\ref{eqn:ploss}). Both channels are quadratically suppressed, but the suppression factor is different, emphasizing the distinction between the underlying physical processes: detuning from the EIT window couples to bright polariton manifolds that live in the resonator and are thus suppressed by both the light-matter-coupling- and control- fields, while inhomogeneous broadening couples to non-resonator bright polariton manifolds that ``see'' only the control- field.

\section{Polariton-polariton Interactions} \label{polaritonpolaritoninteractions}

The interaction between Rydberg atoms will result in an interaction between polaritons, much as in the 1D free-space situation ~\cite{weatherill2008electromagnetically, pritchard2010cooperative, Peyronel2012, Bienias2014}. In the limit that the interaction length-scale is comparable to the mode-waist of the resonator, there can be a substantial renormalization of the collective atomic excitation which we investigate in Sec ~\ref{effectivetheory}. For now, we assume sufficiently weak interactions that photonic- and collective-atomic- components of the polariton wave-function share the same spatial structure; note that these interactions can still dominate over kinetic and potential energies, as well as particle decay rates, so this need not be a ``weakly interacting'' polaritonic gas in the traditional (mean-field) sense.

The bare 3D interaction between two Rydberg atoms takes the form $V(x-x')=\frac{c_r(\theta) C_6}{|x-x'|^6}$ [40], where $c_r(\theta)$ is the angular dependence of the interaction. S-Rydberg atoms have radially symmetric wavefunctions and so $c_r(\theta)\approx c_r \equiv 1$. In the second quantized picture, for a thin atomic cloud (thickness $T \ll d$, where $d$ is the cavity analog of the blockade radius~\cite{Gorshkov2011}) the 2D-projected interaction takes the form (with $ \tilde{V}(x) \approx T \times V(x;z=0)$): 

\begin{eqnarray}
H_{int}=\frac{1}{2} \iint \mathrm{d}x \textrm{ } \mathrm{d}x' \nonumber \\
\phi^{\dagger}_r(x) \phi^{\dagger}_r(x') \tilde{V}(x-x')  \phi_r(x') \phi_r(x). \nonumber
\end{eqnarray}

\noindent $\phi^{\dagger}_r(x)$ may be written in the polariton basis in analogy to the way $a^{\dagger}(x)$ was written in the polariton basis in the preceding section:

\begin{eqnarray}
H_{int}=\frac{1}{2} \sum_{ijkl \in \boldmath{[} d, b_-, b_+ \boldmath{]}} d_{i}d_{j} \tilde{d}_k \tilde{d}_l \nonumber \\
\iint \mathrm{d}x \textrm{ } \mathrm{d}x' \textrm{ } \chi^{\dagger}_i(x)\chi^{\dagger}_j(x') \tilde{V}(x-x') \chi_k(x')\chi_l(x). \label{interactionhamiltonian}
\end{eqnarray}

\noindent Here $d_i,\tilde{d}_j$ are matrix elements of the inverse $\mu,\tilde{\mu}$ matrices: $d_j \equiv \mu^{-1}_{ryd,j},\tilde{d}_j \equiv \tilde{\mu}^{-1}_{ryd,j}$ and the index ``ryd'' denotes the Rydberg slot of $\mu^{-1},\tilde{\mu}^{-1}$. In the absence of Rydberg loss and 2-photon detuning, $d_{d}=\sin \frac{\theta_{d}}{2}$.

If the interaction energy $\tilde{V}$ is small compared to the splitting between dark- and bright- polariton branches (Sec.~\ref{interactiondrivenloss} explores the couplings and loss that violate this condition), the diagonal elements of $H_{int}$ dominate, yielding a lowest-order polariton-projected effective Hamiltonian:

\begin{empheq}[box=\fbox]{align}
\hat{P}_{d}(H_{tot}+H_{int})\hat{P}_{d}= \nonumber \\
\cos^2\frac{\theta_{d}}{2}  \Bigl (\int \mathrm{d}x\chi^{\dagger}_{d} 
\boldmath{[} \frac{\mathbf{\Pi}^2}{2m_{p}} + \frac{1}{2} m_{p} \omega^2_{t}|x|^2-\frac{i \kappa}{2}\boldmath{]} \chi_{d} \Bigr ) \nonumber \\
+\frac{1}{2} \sin^4\frac{\theta_{d}}{2} \left (\iint \mathrm{d}x\mathrm{d}x' \hat{n}_d(x) \hat{n}_d(x')\tilde{V}(x,x') \right ), \label{totalhamiltonian}
\end{empheq}

\noindent where $\hat{P}_{d}$ is dark-polariton projection operator, and we have defined the dark polariton number density operator $\hat{n}_d(x)=\chi^{\dagger}_{d}(x)\chi_{d}(x)$.

By tuning the dark-state rotation angle $\theta_d$ (via atomic density and control-field intensity) it is possible to move from a weakly interacting gas of ``nearly-photonic'' polaritons (for $\theta_d\approx 0$) to a strongly interacting gas of ``nearly-Rydberg'' polaritons (for $\theta_d=\frac{\pi}{2}$) and explore the correlations which then develop~\cite{Sommer2015}. In this latter limit it is likely that the interactions $\tilde{V}$ become comparable to the dark-bright splitting, so the interaction potential must be renormalized as explored in section ~\ref{effectivetheory}.

\section{Interaction Driven Polariton Loss} \label{interactiondrivenloss}

The polariton-projected field theory of the preceding sections neglects loss from collisional coupling to bright manifolds. We investigate processes of this sort by exploring the minimal model of two $\delta$-interacting dark polaritons in the TEM$_{00}$ mode of a resonator, with $\tilde{V}(x,x')=U_{eff}\delta(x-x')$, where $U_{eff}$ is a phenomenological interaction strength. Such an interaction couples the two dark polariton state $|dd \rangle$ to a final two-polariton state $|f \rangle$ with Rabi frequency $\Omega_{dd \rightarrow f}=\langle f|\tilde{V}|dd \rangle$. We consider two final states: (1) one each upper and lower bright polaritons ($|b_+b_- \rangle$), which is energetically degenerate with $|dd \rangle$, and (2) dark- and (upper/lower) bright- polariton ($|db\rangle$), which is off-resonant, but sufficiently spectrally broad that its enhanced matrix element makes it important.

\begin{figure}[h]
     \centering
     \subfloat[][]{\includegraphics[scale=0.3,valign=c]{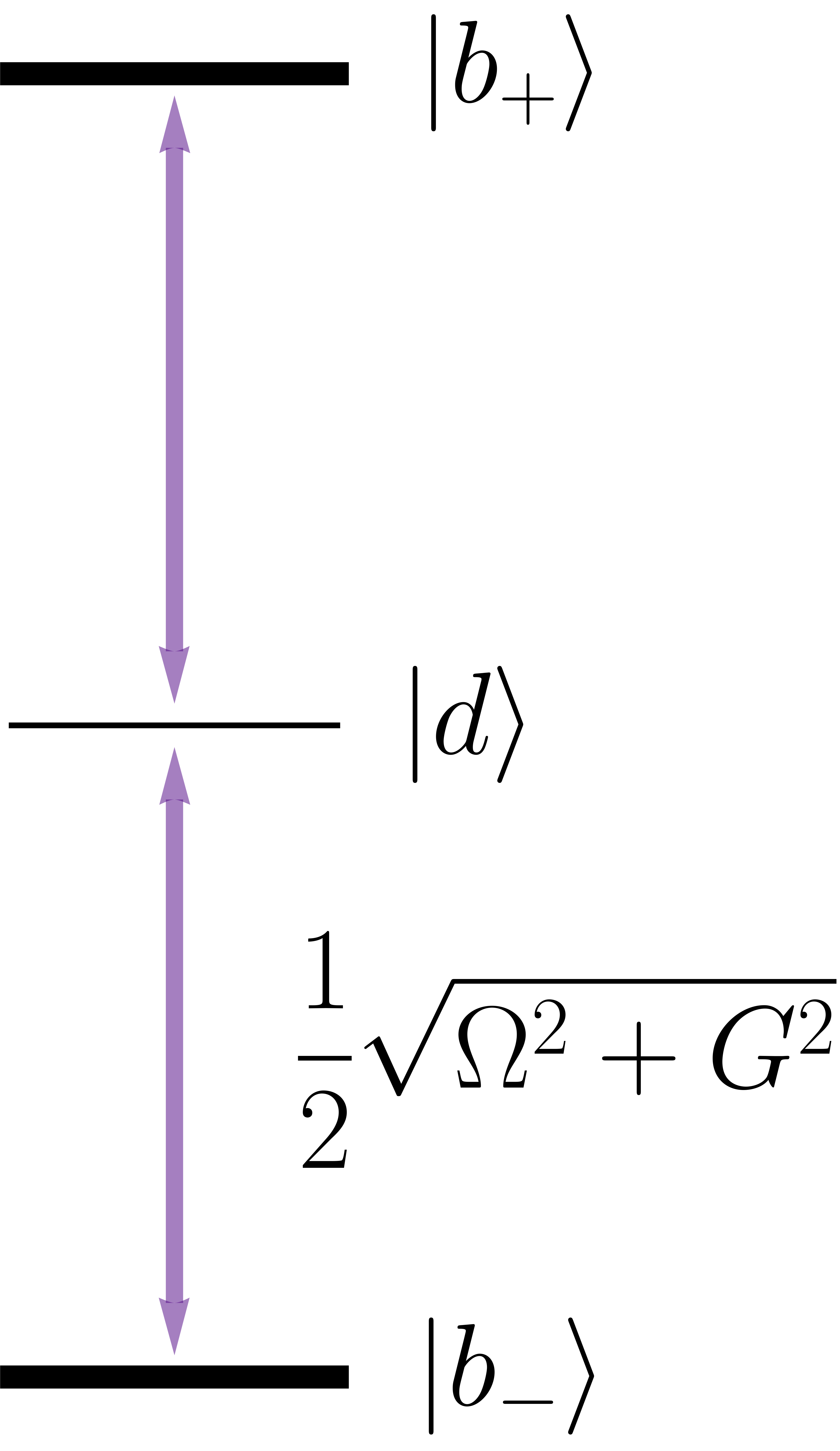}\label{lab1}}
     \subfloat[][]{\includegraphics[scale=0.3,,valign=c]{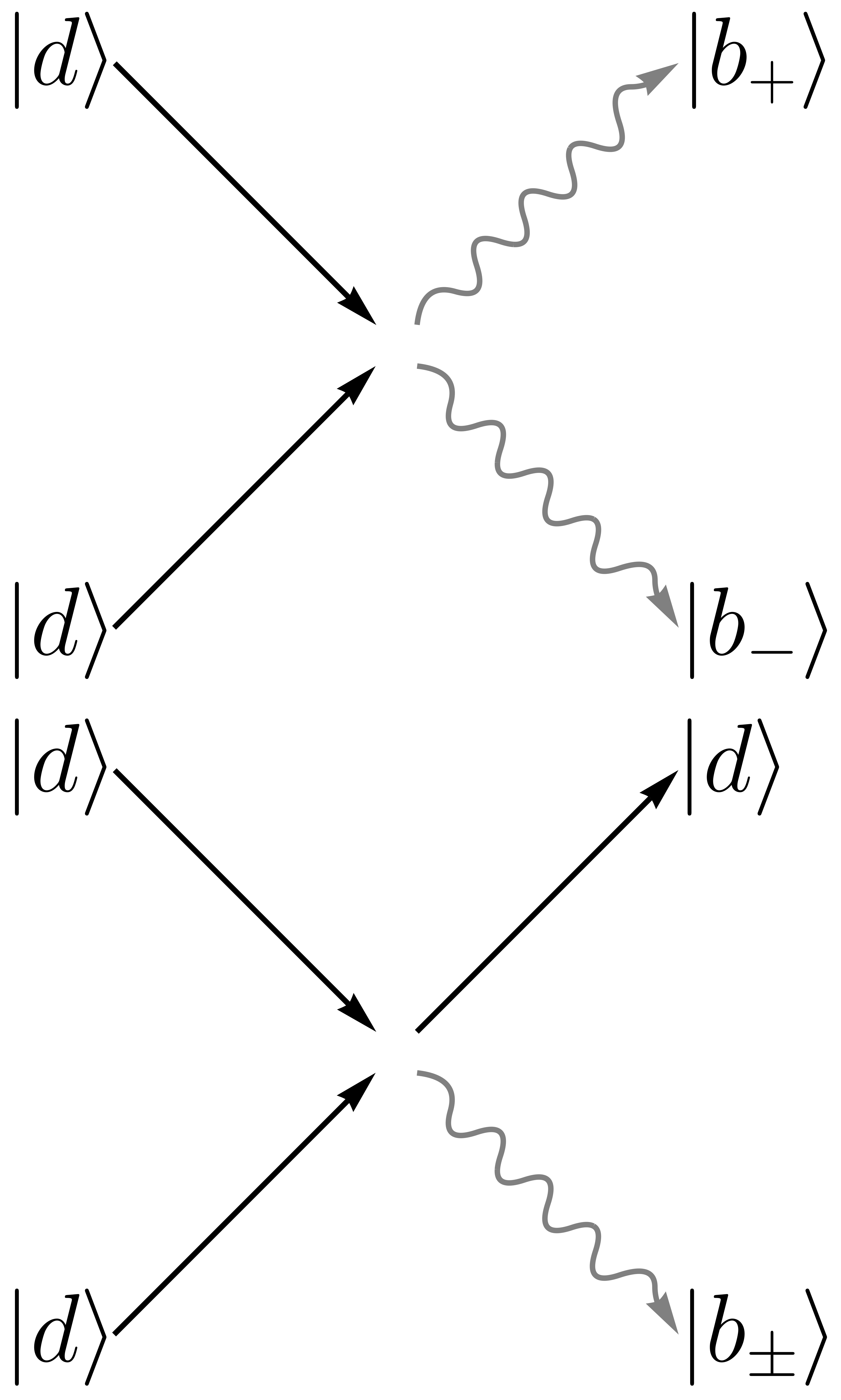}\label{lab3}}
\caption{\textbf{Interaction driven loss. (a)} In the rotating frame, the dark polariton $|d \rangle$ has zero energy.  The upper and lower bright polariton branches ($|b_+ \rangle$ and $|b_- \rangle$ respectively) are separated in energy from the dark polariton by the coupling and control lasers, giving a detuning of $\pm \frac{\sqrt{G^2+\Omega^2}}{2}$. \textbf{(b)} Two dark polaritons that experience a contact interaction $U_{eff}$ can become a pair of bright polaritons $|b_+b_- \rangle$ (upper diagram) or a dark and (upper- or lower- branch) bright polariton $|db_\pm \rangle$ (lower diagram). Dark polaritons are depicted by a straight black line while the wavy gray lines are bright polaritons.  The form of the additional dark polariton loss introduced by these scattering events are determined by how ``resonant'' the processes are: the $\sim$energy conserving $|dd \rangle \rightarrow |b_+b_- \rangle$ process introduces a loss of $\Gamma_{dd \rightarrow bb}=\frac{\sin^4 \theta_d U^2_{eff}}{8\gamma_e}$, while the $|dd \rangle \rightarrow |db_\pm \rangle$ coupling is ``off-resonant'' and so its additional loss $\Gamma_{dd \rightarrow db} = \sin^6 \frac{\theta_d}{2} \sin^2 \theta_d \frac{\gamma_e}{G^2} U^2_{eff}$ is suppressed by the light-matter coupling field.}
\label{fig:darkdarkbrightbrightcouplingfig}
\end{figure}

In a frame rotating with the cavity and control fields ($\delta_e=\delta_2=0$), the dark polaritons have zero energy and the upper/lower bright polariton branches are energetically shifted by $\pm \frac{\sqrt{G^2+\Omega^2}}{2}$ (Fig.~\ref{fig:darkdarkbrightbrightcouplingfig}). The process $|dd\rangle\rightarrow|b_+b_- \rangle$ process is thus energy conserving, with collisional Rabi coupling given by: 

\begin{eqnarray}
\Omega_{dd \rightarrow bb}&=&\langle b_+b_-|\tilde{V}|dd \rangle=\sqrt{2}\sin^2 \frac{\theta_d}{2} \cos^2 \frac{\theta_d}{2} U_{eff} \nonumber \\
&=& \frac{1}{2\sqrt{2}} \sin^2 \theta_d U_{eff}. \nonumber
\end{eqnarray}

\noindent Because this process is resonant, it induces a loss $\Gamma_{dd \rightarrow bb}=\frac{\Omega^2_{bb}}{\Gamma_{bb}}$, where $\Gamma_{bb}=2\frac{\gamma_e}{2}$ is the intrinsic loss of the bright polariton branches which are predominantly composed of the lossy P-state. Plugging in, the loss rate is:

\begin{eqnarray}
\Gamma_{dd \rightarrow bb}=\frac{\sin^4 \theta_d U^2_{eff}}{8\gamma_e}.
\end{eqnarray}

\noindent This loss depends heavily on the interaction strength between the Rydberg atoms and the dark state rotation angle. Making the particles more Rydberg-like creates stronger interactions and increases the loss rate; similarly, using a higher principle quantum number increases $U_{eff}$, further enhancing the loss.

We now investigate the second scattering process, $|dd \rangle \rightarrow |db \rangle$  (upper or lower bright polariton). Following the same procedure, the collisional Rabi frequency is:

\begin{eqnarray}
\Omega_{dd \rightarrow db}&=&\langle db|\tilde{V}|dd \rangle=\sqrt{2}\sin^3 \frac{\theta_d}{2} \cos \frac{\theta_d}{2} U_{eff}. \nonumber
\end{eqnarray}

\noindent Since this is an off-resonant process (final and initial state are detuned by $\Delta=\frac{\sqrt{G^2+\Omega^2}}{2}$), the resulting loss rate is $\Gamma_{dd \rightarrow db}=\frac{\Omega^2_{db}}{\Delta^2}\Gamma_{db}$, with $\Gamma_{db}=\frac{\gamma_e}{2}$ (we ignore resonator- and Rydberg- loss, as bright-state loss is dominated by p-state loss in alkali metal atom Rydberg cQED experiments~\cite{Jia2018}). Combining these effects/approximations yields:

\begin{eqnarray}
\Gamma_{dd \rightarrow db} = \sin^6 \frac{\theta_d}{2} \sin^2 \theta_d \frac{U^2_{eff}}{G^2}\gamma_e.
\end{eqnarray}

\noindent Once again, stronger interactions and a more Rydberg-like character for the polaritons increases the loss from this process but due to the off-resonant nature of this process there is a quadratic suppression from the light-matter coupling which  separates the bright- and dark- polaritons in energy.

We can compare the two loss processes:

\begin{eqnarray}
\frac{\Gamma_{dd \rightarrow bb}}{\Gamma_{dd \rightarrow db}}=\frac{\sin^2 \theta_d}{8\sin^6 \frac{\theta_d}{2}} \frac{G^2}{\gamma_e^2}.
\end{eqnarray}

\noindent In the limit $G \gg \Omega,\gamma_e$ ($\frac{\theta_d}{2} \rightarrow \frac{\pi}{2} \Rightarrow \sin \frac{\theta_d}{2} \rightarrow 1, \sin \theta_d \rightarrow 0$), the $|dd \rangle \rightarrow |db \rangle$ loss channel will dominate, while in the opposite limit, interaction-driven loss is dominated by the $\sim$ energy conserving process.

\section{Effective Theory for Interacting polaritons} \label{effectivetheory}

The simple polariton-projected interacting theory introduced in section ~\ref{polaritonpolaritoninteractions} is an accurate description \emph{only} for polaritons whose interaction energy is less than the EIT linewidth~\cite{Jia2016} for all pairs of atoms comprising the polaritons. As two polaritons approach one another and their wavepackets begin to spatially overlap, some terms in the interaction energy diverge; full numerics (see SI of ~\cite{Jia2018} for details of the approach) reveal that the two-polariton wavefunction is renormalized to suppress such overlap, at the cost of additional (finite) interaction energy, and loss. We now explore the extreme case of this physics: a single-mode optical resonator that is moderately-to-strongly blockaded, to develop a low-dimensional effective model in the basis of near-symmetric collective states that the full numerics.

The ``brute force'' numerical approach that we have previously employed accounts for the  three-level structure of each atom in the atomic-ensemble, and the interactions between atoms. It accurately reproduces observed correlations ~\cite{Jia2018} at the expense of a Hilbert space which grows as $N^m$, where $N$ is the number of atoms in the atomic ensemble and $m$ the number of polaritons in the system; including multiple resonator transverse modes to allow for motional dynamics of the polaritons rapidly becomes computationally intractable. The problem of an extremely large Hilbert space is exacerbated since many-body physics~\cite{Sommer2015,Schine2018} demands both multiple resonator modes and significantly more than two excitations, which makes numerically computing the behavior of the system completely untenable without a coarse-grained effective theory.

\begin{figure*}[ht]
     \centering
     \subfloat[][]{\includegraphics[scale=0.275,valign=c]{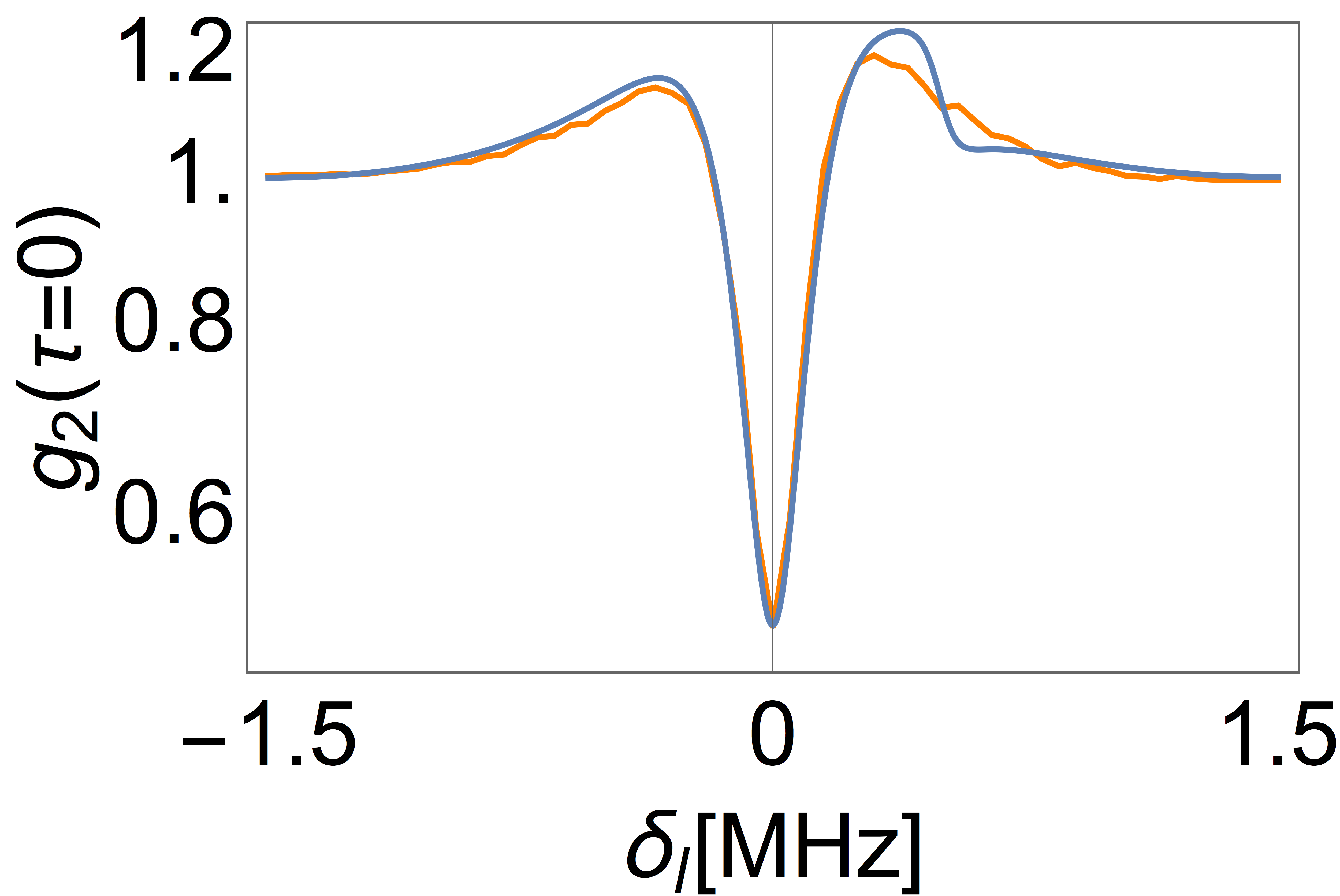}\label{g2deltal}}
     \subfloat[][]{\includegraphics[scale=0.275,valign=c]{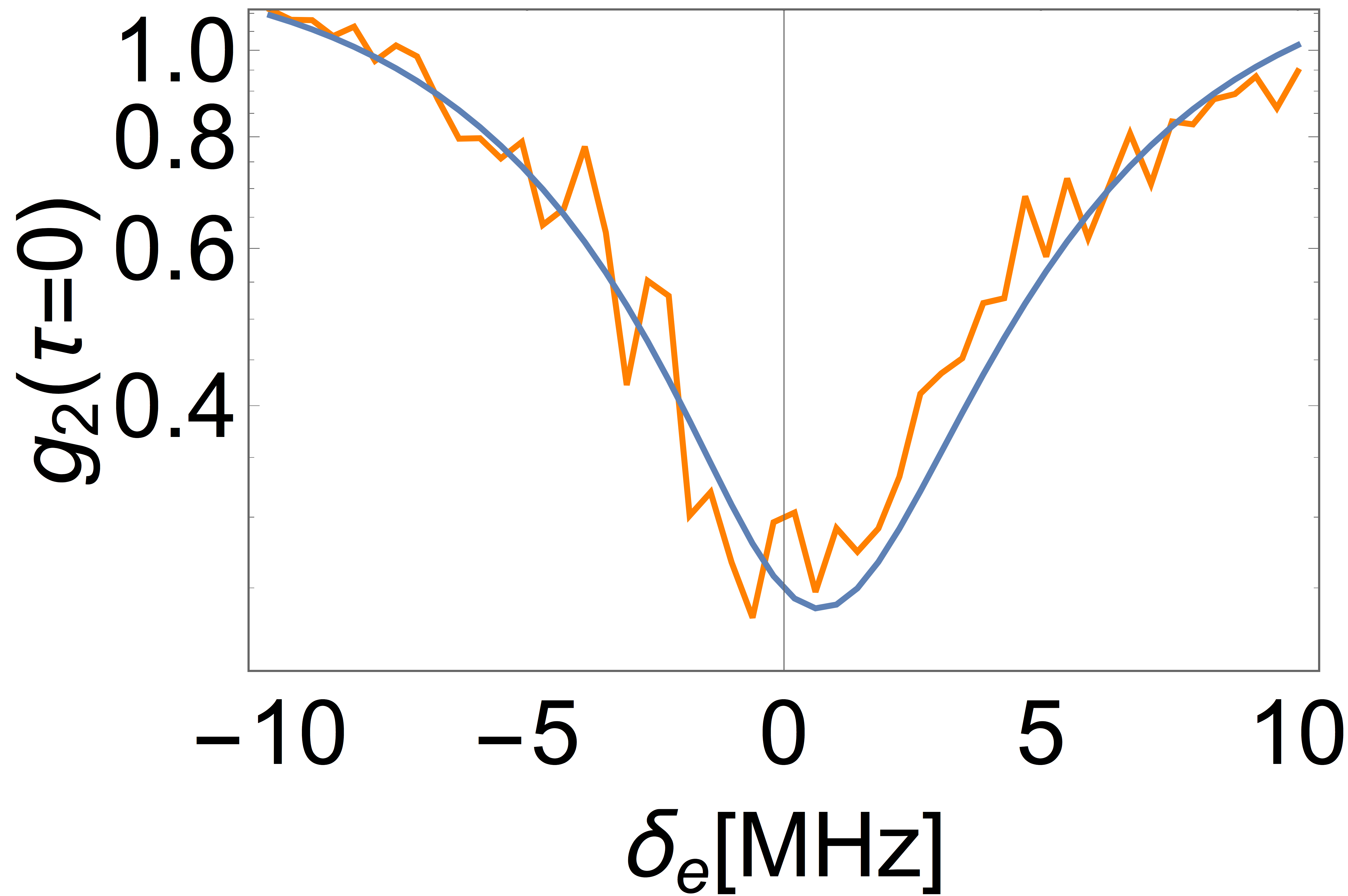}\label{g2deltae}}
     \subfloat[][]{\includegraphics[scale=0.275,valign=c]{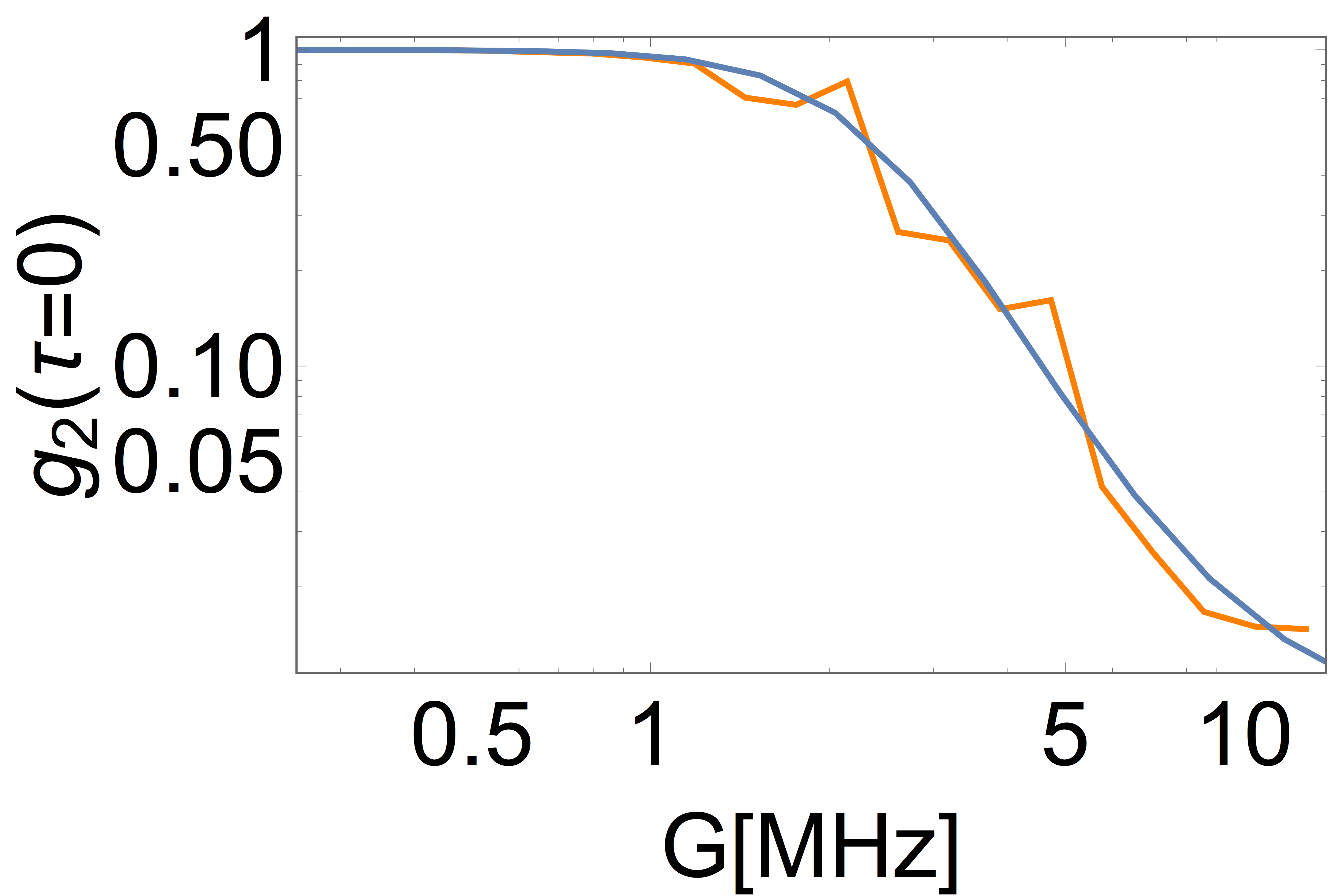}\label{g2G}}
     \subfloat[][]{\includegraphics[scale=0.275,,valign=c]{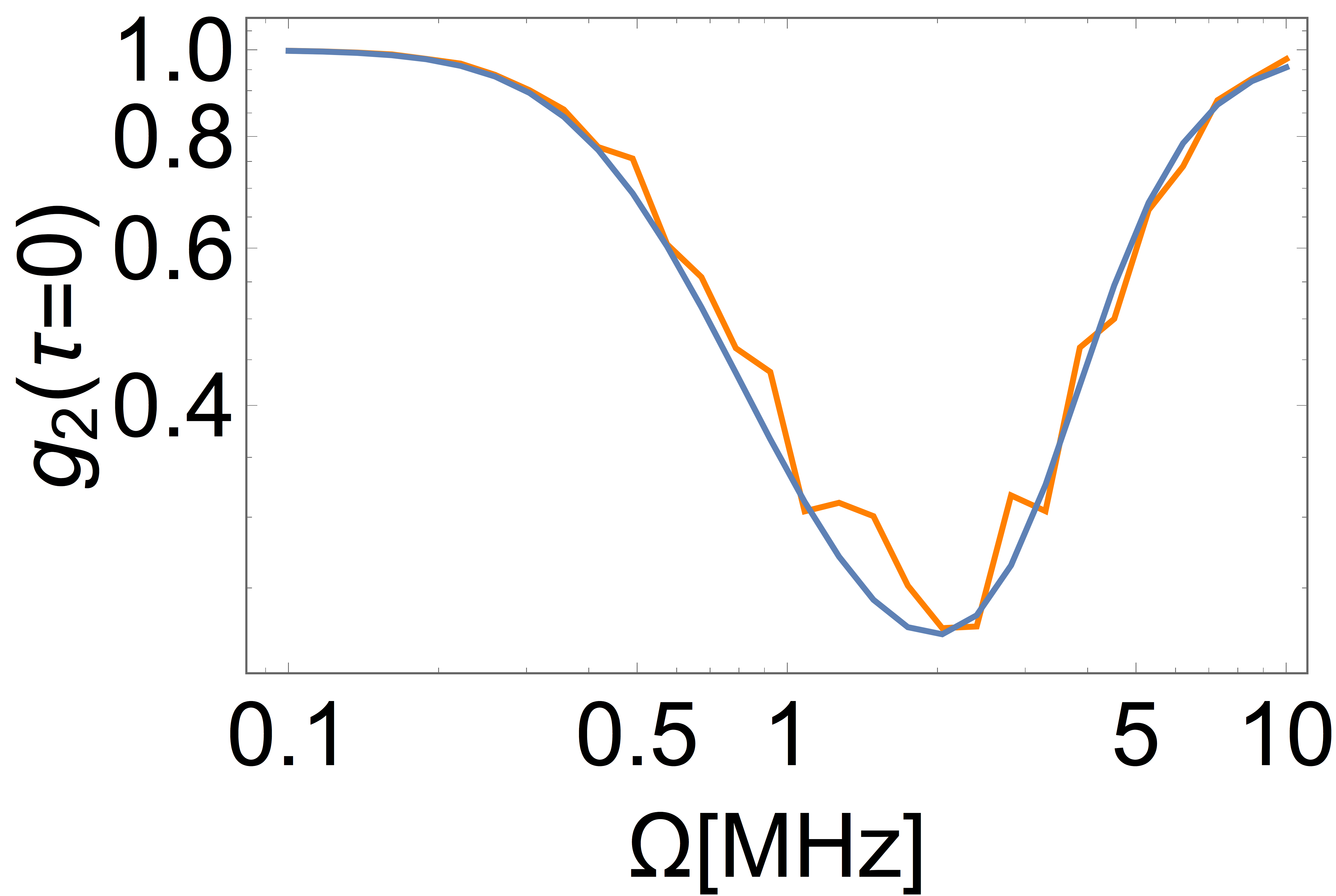}\label{g2Omega}}
     \subfloat[][]{\includegraphics[scale=0.275,,valign=c]{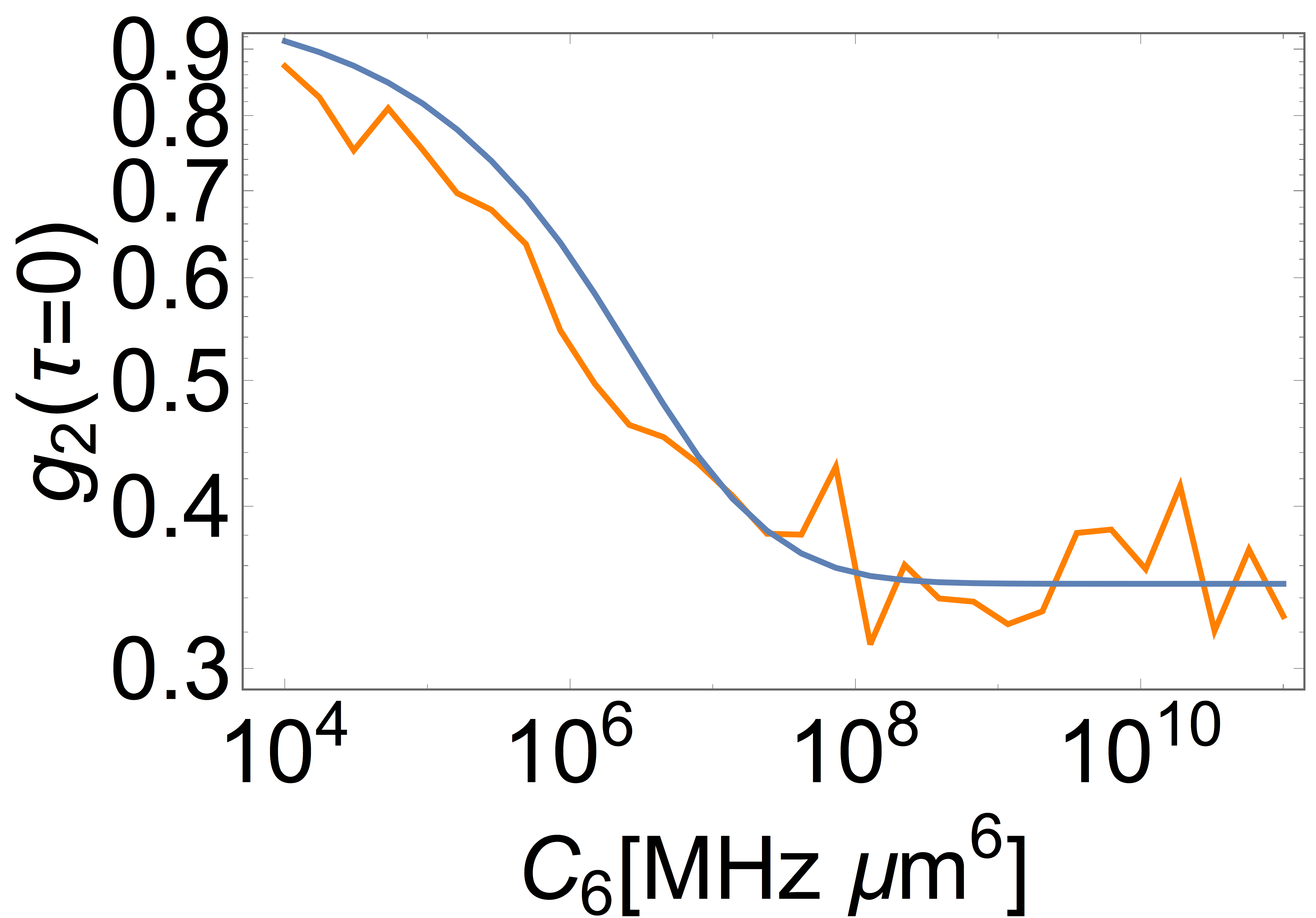}\label{g2C6}}
     \caption{\textbf{Comparing the effective field theory to full numerics.} Effective theory renormalized interactions and couplings (blue line) and full numerical model (orange line) are compared by looking at the temporal intensity autocorrelation function $g_2(\tau=0)$, which characterizes the strength of interactions in the system. Common parameters are (unless varied in the plot or mentioned otherwise): $G=2\pi \times 6$ MHz, $\Omega=2\pi \times 2.3$ MHz, $\delta_e$= 0 MHz, $C_6=2\pi \times 56 $ THz $\mu$m$^6$ (corresponding approximately to the interaction strength of the 100S Rydberg state), $N_{atom}=700$, $\gamma_e=2\pi \times 6$ MHz, $\kappa=2\pi \times$1.6 MHz, $\gamma_r=2\pi \times 0.15$ MHz, $\delta_c=0$ MHz, $w_c=14$ $\mu$m. \textbf{(a)}$g_2(\tau=0)$ vs probe laser detuning. Scanning the probe laser frequency affects both the detuning to the P-state and the two-photon detuning to the Rydberg state. \textbf{(b)} $g_2(\tau=0)$ vs P-state detuning $\delta_e$. The detuning of the P-state $\delta_e$ is varied while keeping the two photon detuning $\delta_2$ zero. \textbf{(c)} $g_2(\tau=0)$ vs light-matter coupling field strength $G$, $g_2(\tau=0)$ vs probe detuning $\delta_l$ (inset). The light-matter coupling strength per atom $g$ is varied from $2\pi \times 0.07$ MHz to $2\pi \times 3$ MHz. For the inset, we scan the probe frequency.\textbf{(d)} $g_2(\tau=0)$ vs control field strength $\Omega$. There is good agreement between the full numerics and the renormalized effective theory except for a small region at about $\Omega = 2\pi \times 1-3$ MHz. \textbf{(e)}$g_2(\tau=0)$ vs Rydberg-Rydberg interaction coefficient $C_6$. $C_6$ is varied from $2\pi \times 10^4$ MHz $\mu$m$^6$ to $2\pi \times 10^{11}$ MHz $\mu$m$^6$, which corresponds to Rydberg states in the range $n\sim$ 48-193S. A control field of $\Omega=2 \pi \times 1$ MHz was employed for this plot. The first-principles renormalized theory accurately reproduces the blockade from weak- to strong- interactions.} \label{numericscomparison}
     \label{g2stuff}
\end{figure*}

To the extent that ``polaritons'' are well-defined collective excitations whose atomic spatial structure reflects the cavity mode functions, it should be possible to develop an effective theory whose Hilbert space size is independent of the atom number, making explorations of multimode/many-body physics tractable. In this section we demonstrate, for a single optical resonator mode, an approach to handle the suppression of short-range double-excitation of the Rydberg-manifold, arriving at a coarse-grained effective theory including both dark and bright polaritons whose parameters may be calculated from first principles.

Consider two excitations, either atomic or photonic, in an atomic ensemble coupled to a single-mode resonator. In the absence of interactions, we can explicitly write out the collective states that couple to the resonator field: $|CC \rangle$, $|CE \rangle$, $|CR \rangle$, $|EE \rangle$, $|ER \rangle$, $|RR \rangle$. These states represent two excitations as photons in the resonator, one photon and one p-state excited atom, one photon and one Rydberg atom, two excited p-state atoms, one p-state and one Rydberg atoms and two Rydberg atoms respectively. The Hamiltonian is closed in this basis, and takes the form~\cite{simon2010cavity}:

\begin{equation*}
\begin{aligned}
H=\left(
\begin{array}{cccccc}
|CC \rangle & |CE \rangle & |CR \rangle & |EE \rangle & |ER \rangle & |RR \rangle \\
\hline
-i \kappa & \frac{G}{\sqrt{2}} & 0 & 0 & 0 & 0  \\
\frac{G}{\sqrt{2}} & -i\frac{\kappa+\tilde{\gamma}_e}{2} & \frac{\Omega}{2} & \frac{G}{\sqrt{2}}& 0 & 0  \\
0 & \frac{\Omega}{2} & -i \frac{\kappa+\gamma_r}{2} & 0 & \frac{G}{2} & 0  \\
0 & \frac{G}{\sqrt{2}} & 0 & -i\tilde{\gamma}_e & \frac{\Omega}{\sqrt{2}} & 0  \\
0 & 0 & \frac{G}{2} & \frac{\Omega}{\sqrt{2}} & -i\frac{\tilde{\gamma}_e+\gamma_r}{2} & \frac{\Omega}{\sqrt{2}}  \\
0 & 0 & 0 & 0 & \frac{\Omega}{\sqrt{2}} & -i \gamma_r 
\end{array}
\right)
\end{aligned}
\end{equation*}

\noindent where $\tilde{\gamma}_e\equiv \gamma_e+2 i \delta_e$ is a complex linewidth incorporating the P-state detuning. The above basis and corresponding Hamiltonian no longer accurately describe the physics once the Rydberg-Rydberg interactions become comparable to the dark-bright splitting: under such conditions, the $|RR \rangle$ is renormalized due to Zeno suppression of excitation of Rydberg-atom-pairs at small separation. We posit that the model can be ``fixed'' by considering coupling to a new collective ``two-Rydberg'' state $|\widetilde{RR} \rangle$, where the tilde signifies that the relative two-Rydberg amplitudes are renormalized by interaction; furthermore, the coupling from $|ER \rangle$ to $|\widetilde{RR} \rangle$ will no longer be $\frac{\Omega}{\sqrt{2}}$.

To ascertain the form of the state $|\widetilde{RR} \rangle$, we will examine the equations of motion \emph{in the frequency domain} under the non-Hermitian Hamiltonian in the bare-atomic basis within the two- excitation manifold. We work in a frame that rotates with an energy $2\Omega_p$, convenient for performing scattering experiments of pairs of photons injected by a probe at energy $\Omega_p$. We assume that while the state $|RR\rangle$ is renormalized by the interactions, the state $|ER\rangle$ is \emph{not}, and reflects the non-interacting polaritonic wave-functions of the preceding sections; this is the central assumption of this section, and is validated by numerics. Corrections to $|ER\rangle$ would enlarge the Hilbert space and may be included as higher-order terms in the effective theory.

Following notation from ~\cite{Jia2018} SI L, the equation of motion for the amplitude of two Rydberg excitations in atoms $\alpha$ and $\beta$, $C_{RR}^{\alpha \beta}$, is given by $i2 \Omega_p C_{RR}^{\alpha \beta}=i\left ( U^{\alpha \beta}_{RR}+2\delta_2 + i \gamma_r \right )C_{RR}^{\alpha \beta}+i\Omega \left ( C_{ER}^{\alpha \beta}+C_{RE}^{\alpha \beta}\right )$, where we have assumed that the control field $\Omega$ is uniform across the atomic ensemble. Here $C_{ER}^{\alpha \beta}$ and $C_{RE}^{\alpha \beta}$ are the amplitudes to have P- and Rydberg- excitations in atoms $\alpha,\beta$ and $\beta,\alpha$, respectively, and satisfy $C_{ER}^{\alpha \beta}=C_{RE}^{\beta \alpha}$. The assumption that $|ER\rangle$ is not renormalized by the interactions is equivalent to $C_{ER}^{\alpha \beta}=\frac{g_\alpha g_\beta}{\sum_\nu |g_\nu|^2}$. Plugging this expression into the equation of motion for $C_{RR}^{\alpha \beta}$ yields the un-normalized two-Rydberg state amplitude: $C_{RR}^{\alpha \beta}=\frac{\Omega \frac{g_\alpha g_\beta}{\sum_\nu |g_\nu|^2}}{\left ( U^{\alpha \beta}_{RR}+2\tilde{\delta}_r \right )}$, where we defined the complex detuning $\tilde{\delta}_r=\delta_2 + i \frac{\gamma_r}{2}-\Omega_p$. We can now write the normalized collective state $|\widetilde{RR} \rangle$, its effective interaction energy $\tilde{U}$ and effective coupling $\frac{\tilde{\Omega}}{\sqrt{2}}$ to $|ER\rangle$ as:

\begin{empheq}[box=\fbox]{align}
|\widetilde{RR} \rangle &=& \frac{\sum_{\alpha\beta}\frac{g_\alpha g_\beta }{ U^{\alpha \beta}_{RR}+2\tilde{\delta}_r }|R_\alpha R_\beta \rangle}{\sqrt{\sum_{\mu\nu}|\frac{g_\mu g_\nu }{ U^{\mu \nu}_{RR}+2\tilde{\delta}_r }|^2}}, \label{collectiveRRstate}\\
\tilde{U} &=& \langle \widetilde{RR} | U | \widetilde{RR} \rangle \nonumber \\
&=&  \frac{\sum_{\alpha\beta}|\frac{g_\alpha g_\beta }{ U^{\alpha \beta}_{RR}+2\tilde{\delta}_r }|^2U_{RR}^{\alpha \beta}}{\sum_{\mu\nu}|\frac{g_\mu g_\nu }{ U^{\mu \nu}_{RR}+2\tilde{\delta}_r }|^2} \label{effectiveinteraction}, \\
\frac{\tilde{\Omega}}{\sqrt{2}} &=& \langle \widetilde{RR}|\sum_j \frac{\Omega}{2} \left ( \sigma_{er}^j+\sigma_{re}^j \right ) | ER \rangle \nonumber \\
&=& \frac{\frac{1}{\sqrt{2}}\Omega  \sum_{\alpha\beta}\frac{|g_\alpha g_\beta|^2 }{ U^{\alpha \beta}_{RR}+2\tilde{\delta}_r }}{\sqrt{\sum_{\mu\nu}|\frac{g_\mu g_\nu }{ U^{\mu \nu}_{RR}+2\tilde{\delta}_r }|^2}\sqrt{\sum_{\mu\nu}|g_\mu g_\nu|^2}}. \label{effectiveOmega}
\end{empheq}

In the extreme limit of strong interactions across all space $U_{RR}^{\alpha \beta} \gg \tilde{\delta}_r$: $\tilde{U}= \frac{\sum_{\alpha\beta}\frac{|g_\alpha g_\beta|^2 }{ U^{\alpha \beta}_{RR}}}{\sum_{\mu\nu}\frac{|g_\mu g_\nu|^2 }{ \left (U^{\mu \nu}_{RR} \right )^2}}=\frac{C_6}{w_c^6}\frac{\int \mathrm{d}\tilde{A}\mathrm{d}\tilde{A}'e^{-2(\tilde{r}^2+\tilde{r}^{'2})}\tilde{d}^6}{\int \mathrm{d}\tilde{A}\mathrm{d}\tilde{A}'e^{-2(\tilde{r}^2+\tilde{r}^{'2})}\tilde{d}^{12}}\approx \frac{1}{120} \frac{C_6}{w_c^6}$, where $w_c$ is the mode waist; the pre-factor makes this interaction substantially weaker than one might na\"ively anticipate- the  interaction predominantly arises from particles separated by $\sim 2.2 w_c$, and not $w_c$.

We next benchmark the validity of this effective theory against a full microscopic numerical model ~\cite{Jia2018}. As a figure of merit we have chosen the temporal intensity autocorrelation function $g_2(\tau)$, which compares the rate at which pairs of photons escape the resonator with separation in time of $\tau$ to what would be expected for uncorrelated photons escaping at the same average rate; in an experiment where our only access to the Rydberg physics is through photons leaking from the resonator, $g_2(\tau)$ characterizes the strength of polariton interactions, with $g_2(\tau=0)\ll 1$ indicating strong interactions between intracavity photons. We compare $g_2(0)$ vs. probe detuning $\delta_l$ (Fig.~\ref{g2deltal}), $g_2(0)$ vs. p-state detuning $\delta_e$ (Fig.~\ref{g2deltae}), light-matter coupling strength $G$ (Fig.~\ref{g2G}), Rydberg control field strength $\Omega$ (Fig.~\ref{g2Omega}) and van der Waals interaction coefficient $C_6$ (Fig.~\ref{g2C6}) between brute-force numerics of many individual three-level atoms and the effective theory developed above. It is apparent that our approach largely agrees with the full numerical model up to ``noise'' arising from randomness in the atom locations. We expect that residual deviations can be parameterized as corrections to $\tilde{\Omega}$ and $\tilde{U}$ due to coupling to bright-polariton manifolds, and a slight enlargement of the Hilbert space to incorporate the additional states coupled to.

\section{Momentum-Space Interactions} \label{offplaneinteractions}

A resonator which exhibits manifolds of nearly-degenerate modes may be understood as a self-imaging cavity: a localized spot living within such a manifold is re-focused onto itself after a full transit around the cavity. In-between, the localized spot undergoes diffraction, equivalent to the time-of-flight expansion of a free atomic gas~\cite{ketterle1999making}. Indeed, what the optics community calls a ``fourier plane'' is what a cold-atom experimentalist calls ``momentum space'': the momentum of the photon in the initial (``reference'' or ``image'') plane has been mapped onto its position in the ``fourier plane''~\cite{goodman2005introduction}. Accordingly, it should be possible to realize interactions which are local in momentum-space by placing a Rydberg-dressed atomic-ensemble that mediates these interactions in a fourier- or nearly-fourier- plane of the optical resonator.

We explore this idea formally by extending the cavity Floquet Hamiltonian engineering tools of our prior work~\cite{Sommer2016} to the interacting regime. A thin gas of Rydberg-dressed atoms placed in a plane separated from the ``reference''/``image'' plane by a ray-propagation matrix $M = (\begin{smallmatrix} \mathbf{a}&\mathbf{b}\\ \mathbf{c}&\mathbf{d} \end{smallmatrix})$ produces interactions of the form:

\begin{align}
H_{int}=\frac{1}{2} \sin^4\frac{\theta_{d}}{2} \iint \mathrm{d}x \textrm{ } \mathrm{d}x'  \chi^{\dagger}_{d}(x)\chi^{\dagger}_{d}(x') \nonumber \\
\tilde{V} \left (\mathbf{a}(x-x')-\frac{\mathbf{b}}{\hbar k}(\hat{p}-\hat{p}') \right ) \chi_{d}(x')\chi_{d}(x). \label{generalizedinteractionhamiltonian}
\end{align}

For the simple case of a delta-interacting gas of atoms placed in such an intermediate plane (a distance $z$ from the reference plane), we employ this result to transform an expression where the dark polariton creation/destruction operators and interaction potential are written in the intermediate plane to one where the interaction is transformed and all operators are written in the ``reference'' plane:

\begin{align}
H_{int}=\frac{1}{2} \sin^4\frac{\theta_{d}}{2} \iint \mathrm{d}x \textrm{ } \mathrm{d}x' \textrm{ } \chi^{\dagger}_{d}(x;z)\chi^{\dagger}_{d}(x';z) \nonumber \\
\delta(x-x') \chi_{d}(x';z)\chi_{d}(x;z) \nonumber \\
=\frac{1}{2} \sin^4\frac{\theta_{d}}{2} \iiint \mathrm{d}x \textrm{ }\mathrm{d}\delta \textrm{ } \mathrm{d}\Delta \textrm{ } \chi^{\dagger}_{d}(x+\delta)\chi^{\dagger}_{d}(x-\delta) \nonumber \\
 e^{i\frac{k}{2z} \left (\delta^2-\Delta^2 \right )} \chi_{d}(x+\Delta)\chi_{d}(x-\Delta). \nonumber
\end{align}

\noindent The resulting polariton interaction is no longer purely local in real-space, and indeed can ``instantaneously'' transport polaritons through space.

The most extreme example of such an interaction occurs if the mediating gas is placed in a fourier plane of the system, $a=0, b=f$ 

\begin{align}
H_{int}=\frac{1}{2} \sin^4\frac{\theta_{d}}{2} \iint \mathrm{d}x \textrm{ } \mathrm{d}x'  \textrm{ }\chi^{\dagger}_{d}(x)\chi^{\dagger}_{d}(x') \nonumber \\
\tilde{V} \left (-\frac{f}{\hbar k} \left (\hat{p}-\hat{p}' \right ) \right )\chi_{d}(x')\chi_{d}(x), \nonumber
\end{align}

\noindent an interaction that is local in momentum-space.

\section{Impact of Atomic Motion} \label{dopplerdecoherence}

In this section we investigate the effects of atomic motion on the coherence properties of individual Rydberg polaritons in both single- and multi-mode regimes. We relax the assumption, employed to this point in the manuscript, that the atoms remain spatially fixed, and instead allow them to move ballistically through space. The impact of this motion upon the P-state is ignored because the P-state linewidth of an alkali-metal atom ($\sim 2\pi\times 6$MHz) is typically much larger than any Doppler broadening effect at $\mu$K temperatures ($\sim 2\pi\times 100$kHz for Rb), and furthermore, dark polaritons by construction spend very little time in the P-state (they are ``dark'' to it). In what follows, we will assume the polariton is almost entirely Rydberg-like (the typical experimental situation~\cite{Jia2018}; if this is not the case, all doppler-induced broadenings and cross-couplings must be multiplied by a factor reflecting the Rydberg-participation of a polariton $\sin^2{\frac{\theta_d}{2}}$.

We incorporate atomic motion into the Hamiltonian in the bare-atom basis by allowing each atom to have a time-dependent coupling-phase to the probe and control fields resulting from its time-varying position:

\begin{eqnarray}
&H&=\omega_c \hat{a}^\dagger \hat{a} |c\rangle \langle c| + \sum_j \omega_e |e\rangle \langle e| + \omega_r |r\rangle \langle r|\nonumber \\
&+&\sum_j \Big\{ \Big[ G_j \left ( x_j+v_j t \right ) |e\rangle \langle c| + \Omega_j\left ( x_j+v_j t \right ) |r\rangle \langle e| \Big] + \mathrm{h.c.}\Big\},     \nonumber
\end{eqnarray}

\noindent where $x_j$ and $v_j$ are the positions and velocities of the atoms, drawn from a normal distribution reflecting the sample r.m.s. size and temperature. The effect of atomic motion, then, is to mix the collective states that couple to the resonator modes with those that, in the absence of atomic motion, do not couple to it. To see this formally, we write the Hamiltonian in the basis of the instantaneous collective eigenstates, resulting in a Hamiltonian of the form:

\begin{eqnarray}
H=\sum_m E_m(t) |m(t) \rangle \langle m(t) |+ |\dot{r}(t) \rangle \langle r(t)|, \label{timedependenthamiltonian}
\end{eqnarray}

\noindent where $|m(t) \rangle, E_m(t)$ are the instantaneous polaritonic eigenstates and their corresponding energies, and $|\dot{r}(t) \rangle \langle r(t)|$ is an extra term introduced by this time-dependent change of basis, capturing the effects of atomic motion in the instantaneous collective Rydberg state $|r(t)\rangle$. In what follows, we examine the form of this final term for the particular case of twisted resonators which produce a Landau level for light~\cite{Schine2016,Schine2018}, so the mode functions are Laguerre-Gauss $\Psi_l(z\equiv x+iy)=\sqrt{\frac{2^{l+1}}{\pi l!}}z^l e^{-|z|^2}$, with angular momentum $L=l\hbar$. For a polariton in a mode with angular momentum $l\hbar$, $|r(t)\rangle= |r_l(t)\rangle$, with $|r_l(t) \rangle=\frac{\sum_j e^{i \vec{k} \cdot (\vec{x}_j+\vec{v}_j t)}\Psi_l \left (\frac{\vec{x}_j+\vec{v}_j t}{w_c} \right )}{\alpha_l(t)}|j\rangle$, where $|j\rangle$ is the state where all atoms are in the ground state except for the j$^{th}$ which is in the Rydberg state, and $\alpha_l(t)=\sqrt{\sum_j |\Psi_l \left (\frac{x_j+v_j t}{w_c} \right )|^2}$ is the normalization factor for mode $l$. The time derivative $\frac{d}{dt}|r_l(t) \rangle$ is:

\begin{eqnarray}
|\dot{r}_l(t)\rangle&=&\frac{\sum_j \vec{k} \cdot \vec{v}_j e^{i \vec{k} \cdot \vec{r}}\Psi_l \left (\frac{\vec{r}}{w_c} \right )}{\alpha_l(t)}|j\rangle \nonumber \\
&+&|r_l(t)\rangle \left (-\frac{\dot{\alpha}_l(t)}{\alpha_l(t)} \right ) \label{rdot} \\
&+&\frac{\sum_j \frac{e^{i \vec{k} \cdot \vec{r}}}{w_c} \left (\vec{v}_j \cdot \vec{\nabla}_{\vec{r}} \right )\Psi_l \left (\frac{\vec{r}}{w_c} \right )}{\alpha_l(t)}|j\rangle, \nonumber 
\end{eqnarray}

\noindent where the index $|j \rangle$ runs over all atoms in the sample, $\vec{r}=\vec{x}+\vec{v} t$, $w_c$ is the resonator mode waist, $k$ is the wavevector defined by the relative orientation of the cavity- and control- fields  and $\vec{\nabla}_{\vec{r}}$ refers to the gradient with respect to $\vec{r}$. These terms of $H$ in the instantaneous eigen-basis have three effects: mixing polaritons in modes of different angular momenta, coupling to bright polariton manifolds orthogonal to the resonator field, and random shifts of the energy of the mode in which the polariton resides. We now investigate the extent to which each of the terms above induce each of these effects. Define: 

\begin{eqnarray}
|T_1 \rangle= \frac{\sum_j \vec{k} \cdot \vec{v}_j e^{i \vec{k} \cdot \vec{r}_j}\Psi_l \left (\frac{\vec{r}_j}{w_c} \right )}{\alpha_l(t)}|j\rangle. \nonumber 
\end{eqnarray}

\noindent Even for a maximally degenerate concentric cavity, most collective Rydberg states that one can generate (for example through atomic motion, above) are orthogonal to all resonator modes, because their spatial form \emph{along} the cavity axis does not match the cavity field (equivalently, their longitudinal momentum is not that of a cavity photon). As a consequence, most of the dynamics generated by coupling to $|T_1\rangle$, $|T_2\rangle$, and $|T_3\rangle$ consists of coupling to bright polariton manifolds with no corresponding dark (resonator-like) mode. We bound these effects by assuming, at zeroth order, that \emph{all} of each coupling is to these bright manifolds. The strength of this coupling is thus the normalization of the corresponding $|T_i\rangle$:

\begin{eqnarray}
\langle T_1|T_1 \rangle = \frac{\sum_j k^2 v^2_j |\Psi_l|^2}{\alpha_l^2}, \nonumber \\
\langle T_1|T_1 \rangle^{v}_{t=0}=(k v_{th})^2 \frac{\sum_j  |\Psi_l|^2}{\alpha_l^2} = (k v_{th})^2. \nonumber
\end{eqnarray}

\noindent We can now write $|T_1 \rangle = k v_{th} \frac{|T_1 \rangle}{k v_{th}}=k v_{th} |\tilde{T}_1 \rangle$, where $|\tilde{T}_i \rangle$ is the normalized state-vector corresponding to state $|T_i\rangle$. This corresponds to a Rabi-coupling of strength $\sim k v_{th}$ to a bright polaritonic state which is detuned by $\Omega$, and a resulting dark$\rightarrow$bright loss rate of:

\begin{eqnarray}
\delta \Gamma_{T_1} = \frac{ \left (k v_{th} \right )^2}{\left (\frac{\Omega}{2} \right )^2} \frac{\gamma_e}{2}=\frac{2\left (k v_{th} \right )^2}{\Omega^2} \gamma_e. \label{dopplershift1}
\end{eqnarray}

A small fraction of $|T_1\rangle$ overlaps with other degenerate resonator modes, corresponding to an atomic-motion-induced polaritonic motional diffusion:

\begin{eqnarray}
\langle r_m(t)|T_1 \rangle
= \frac{\sum_p k v_p \Psi_m^* \left (\frac{r_p}{w_c} \right ) \Psi_l \left (\frac{r_p}{w_c} \right )}{\sqrt{\sum_\mu |\Psi_l \left (\frac{r_\mu}{w_c} \right )|^2\sum_\nu |\Psi_m \left (\frac{r_\nu}{w_c} \right )|^2}}. \nonumber
\end{eqnarray}

\noindent The expected value of this term is zero since the average atomic velocity is zero: $\langle v_p\rangle_{t=0}=0$. The r.m.s. coupling, however, is non-zero:

\begin{eqnarray}
\sqrt{\langle |\langle r_m|T_1 \rangle|^2 \rangle ^{v.}_{p.}} &=& \frac{k v_{th}}{\sqrt{N_0}} C'_{l \rightarrow m}, \label{dopplerbroadening1}
\end{eqnarray}

\noindent where $N_0$ is the number of atoms in mode $l=$0 and $C'_{l \rightarrow m}=\frac{C_{l \rightarrow m}}{\sqrt{\sqrt{lm}}}=\sqrt{\frac{2^{1-l-m}\frac{(l+m)!}{l!m!}}{\sqrt{lm}}}$ is a generalized the Doppler coupling matrix element between modes $l$ and $m$, incorporating the fact that higher angular momentum modes contain more atoms, and thus provide a smoother atom distribution. We can expand this matrix element for large $l\approx m$ yielding $C_{l \rightarrow m}\approx e^{-(l-m)^2/2l^2}\frac{\sqrt{2}}{(\pi l)^{1/4}}$, indicating diffusion only into nearly-adjacent modes.

Last, the r.m.s. energy shift (inhomogeneous broadening) of the collective Rydberg state induced by this term is given by:

\begin{eqnarray}
\sqrt{\langle |\langle r_l|T_1 \rangle|^2 \rangle ^{v.}_{p.}} &=& \frac{k v_{th}}{\sqrt{N_0}} \sqrt{\frac{2^{1-2l}(2l)!}{l (l!)^2}}.
\end{eqnarray}

The second term of eq.~\ref{rdot} is:

\begin{eqnarray}
|T_2 \rangle = |r_l \rangle \frac{1}{2w_c} \frac{\sum_n \vec{v}_n \cdot \vec{\nabla}_{\vec{r}}|\Psi_l|^2} {\alpha_l^2}= |r_l \rangle \left ( -\frac{\dot{\alpha}_l}{\alpha_l}\right ). \nonumber
\end{eqnarray}

\noindent Again, we examine how this term couples a dark polariton to the lossy manifold of bright polaritons:

\begin{eqnarray}
\sqrt{\langle T_2|T_2 \rangle} = \frac{v_{th}}{2w_c} \frac{\Gamma(l+\frac{1}{2})}{l!}\approx \frac{v_{th}}{2\sqrt{l+\frac{1}{3}}w_c} . \label{eq:term2broadening}
\end{eqnarray}

\noindent This broadening comes from the time-dependent probe field coupling that the atoms experience as they move within the mode; it is much smaller than $k v_{th}$. From the functional form of $|T_2 \rangle$ we can also see that it does not couple modes of different angular momenta $\left ( \langle r_m|T_2 \rangle = \delta_{l,m} \left ( -\frac{\dot{\alpha}_l}{\alpha_l}\right ) \right )$.

The third term, similar to the first, produces both a broadening and a shift in the dark polariton energy. We can see that this term couples to the bright collective manifold with matrix element:

\begin{eqnarray}
\sqrt{\langle T_3|T_3 \rangle} &=& \frac{v_{th}}{w_c} \sqrt{3l+1}. \label{eq:term3broadening}
\end{eqnarray}

\noindent We can similarly evaluate how this term couples to other states in the dark collective-state manifold:

\begin{eqnarray}
&\langle |\langle r_m|T_3 \rangle |^2 \rangle ^v_p & = \frac{v_{th}^2}{w_c^2}\frac{1}{N_0}  \left (C''_{l \rightarrow m} \right )^2,  \label{dopplerbroadening2}
\end{eqnarray}

\noindent where $C''_{l \rightarrow m}=\sqrt{\frac{2^{-l-m}\left ( l+9l^2+m+2l m+m^2 \right )\Gamma(l+m)}{l!m!\sqrt{lm}}}$ is the coupling element between modes of the resonator that captures mode spatial overlaps, coupling induced by atomic motion and increasing mode area. This cross-thermalization coupling element converges for large $l\approx m$ to $C''_{l \rightarrow m}\approx \sqrt{\frac{6}{\sqrt{\pi}}}l^{-\frac{1}{4}}e^{-(l-m-\frac{11}{6})^2/8l}$. The $l$ dependence in both the broadening and energy shift terms arises from the more rapid phase accrual of higher angular momentum modes.

The total broadening of a mode with angular momentum $l$ is thus $\Gamma_l^{Doppler}\approx 2\frac{(k v_{th})^2}{\Omega^2}\gamma_e\left(1+\frac{3l+1}{(k w_c)^2}\right)$; the r.m.s. Rabi-coupling of mode $l$ to mode $m$ is $|\Omega_{lm}|\approx(\frac{4}{\pi})^{\frac{1}{4}}\frac{k v_{th}}{\sqrt{N_0 l_0}}e^{-\delta l^2/2l_0}\sqrt{1+\frac{\sqrt{9l_0}}{(k w_c)^2}}$, where $2\delta l\equiv l-m$, $2l_0\equiv l+m$. Noting that $k w_c\approx 100$ for typical experiments~\cite{Jia2018}, approximating further yields: $\Gamma_l^{Doppler}\approx 2\frac{(k v_{th})^2}{\Omega^2}\gamma_e$, $|\Omega_{lm}|\approx(\frac{4}{\pi})^{\frac{1}{4}}\frac{k v_{th}}{\sqrt{N_0 l_0}}e^{-\delta l^2/2l_0}$. In fact, summing over all final states, the net coupling out of mode $l$ is $|\Omega_{l}|\approx(16\pi)^{1/4} \frac{k v_{th}}{\sqrt{N_0}}$, independent of $l$.

To summarize, atomic motion results in homogeneous and inhomogeneous broadening of the dark polaritons, along with state diffusion. All effects arise from recoil-induced differential motion of the Rydberg-excited atom, expressed as $\dot{r}_l(t)$: the first couples (eqn's ~\ref{dopplershift1},~\ref{eq:term2broadening},~\ref{eq:term3broadening}) the system to modes that decay into free space due to their spatial symmetry, while the second term  (eqn's ~\ref{dopplerbroadening1}, ~\ref{dopplerbroadening2}) quantifies thermalization of atomic degrees of freedom into the polaritonic degrees of freedom as a result of atomic motion. The former effect is suppressed by the detuning of the uncoupled (and therefore bright) modes from the dark manifold, while the latter effect is suppressed because atomic motion is random, so it is only the shot-noise in the motion of the ensemble comprising the polariton that leads to polaritonic mode coupling.

Doppler decoherence fundamentally arises from the relative motion of the atoms comprising the matter-component of a polariton relative to the field comprising the photonic component; as a consequence, the Doppler decoherence is sensitive to the \emph{canonical} momentum of the optical field, not its \emph{mechanical momentum}~\cite{Sommer2016}. This distinction is particularly important in cavities whose near-degenerate manifolds represent a particle in a magnetic field, because although the Landau level is translationally invariant in a fundamental sense, the choice of gauge arising from resonator twist means that polaritons further from the resonator axis are more susceptible to Doppler decoherence, apparent in the $l$-dependence of the loss terms above.

\begin{table}
\centering
 \begin{tabular}{|c|c|c|} 
 \hline
 \multicolumn{3}{|c|}{Doppler Decoherence Summary} \\
 \hline 
  & Broadening & Cross thermalization \\
 \hline
 $T_1$ & $2\frac{\left ( k v_{th}\right )^2}{\Omega^2}\gamma_e$ & $(\frac{4}{\pi})^{\frac{1}{4}}\frac{k v_{th}}{\sqrt{N_0 l_0}}e^{-\delta l^2/2l_0}$\\
 \hline
 $T_2$ & $\frac{3}{2(3l+1)}\frac{ \left (v_{th}/w_c\right )^2}{\Omega^2}\gamma_e$ & 0  \\
 \hline
 $T_3$ & $2(3l+1)\frac{\left (v_{th}/w_c\right )^2}{\Omega^2}\gamma_e$ & $(\frac{4}{\pi})^{\frac{1}{4}}\frac{v_{th}}{w_c \sqrt{N_0 l_0}}(9 l_0)^{\frac{1}{4}}e^{-(\delta l-\frac{11}{12})^2/2l_0}$\\
 \hline
\end{tabular}
\caption{Atomic motion induced homogeneous/inhomogeneous broadening, as well as r.m.s. diffusion/cross-thermalization matrix elements. For broadening terms, the angular momentum of the state under consideration is $l$; for cross-thermalization we consider  nearby angular momentum states with mean $l_0$ and separation $2\delta l$. Computed broadening and cross-thermalization terms are upper bounds on the respective processes. Note that $k w_c\approx 100$ for typical experiments~\cite{Jia2018}, so $k v_{th}$ terms dominate strongly over $v_{th}/w_c$ terms until $l$ becomes large.}
\end{table}

\section{Outlook} \label{outlook}

In this paper we have presented a field theory of interacting cavity polaritons in the strongly interacting regime, including a formal treatment of interaction and atomic-motion-induced loss channels, and the development of a renormalized single-mode theory. We also demonstrate that by varying the location of one or more Rydberg-dressed atomic ensembles within the resonator, the interactions can be tuned continuously from local in position-space to local in momentum-space.

The renormalized single-mode theory suggests that it should be possible to develop a renormalized cavity-Rydberg-polariton field-theory, analogous to its free-space counterpart~\cite{Gullans2016}, and in conjunction with recently demonstrated cavity Rydberg polariton Keldysh-techniques~\cite{grankin2018diagrammatic}, we are now in a position to accurately model the physics of cavity polariton crystals and Laughlin puddles~\cite{Sommer2015}, plus quantitative analysis of photonic QIP and quantum repeater protocols~\cite{han2010quantum,das2016photonic}.

\section*{Acknowledgements}
The authors would like to thank Nathan Schine for stimulating conversations. This work was supported primarily by MURI grant FA9550-16-1-0323. This work was also supported by the University of Chicago Materials Research Science and Engineering Center, which is funded by National Science Foundation under award number DMR-1420709.

\bibliography{library.bib}

\begin{thebibliography}{51}%
\makeatletter
\providecommand \@ifxundefined [1]{%
 \@ifx{#1\undefined}
}%
\providecommand \@ifnum [1]{%
 \ifnum #1\expandafter \@firstoftwo
 \else \expandafter \@secondoftwo
 \fi
}%
\providecommand \@ifx [1]{%
 \ifx #1\expandafter \@firstoftwo
 \else \expandafter \@secondoftwo
 \fi
}%
\providecommand \natexlab [1]{#1}%
\providecommand \enquote  [1]{``#1''}%
\providecommand \bibnamefont  [1]{#1}%
\providecommand \bibfnamefont [1]{#1}%
\providecommand \citenamefont [1]{#1}%
\providecommand \href@noop [0]{\@secondoftwo}%
\providecommand \href [0]{\begingroup \@sanitize@url \@href}%
\providecommand \@href[1]{\@@startlink{#1}\@@href}%
\providecommand \@@href[1]{\endgroup#1\@@endlink}%
\providecommand \@sanitize@url [0]{\catcode `\\12\catcode `\$12\catcode
  `\&12\catcode `\#12\catcode `\^12\catcode `\_12\catcode `\%12\relax}%
\providecommand \@@startlink[1]{}%
\providecommand \@@endlink[0]{}%
\providecommand \url  [0]{\begingroup\@sanitize@url \@url }%
\providecommand \@url [1]{\endgroup\@href {#1}{\urlprefix }}%
\providecommand \urlprefix  [0]{URL }%
\providecommand \Eprint [0]{\href }%
\providecommand \doibase [0]{http://dx.doi.org/}%
\providecommand \selectlanguage [0]{\@gobble}%
\providecommand \bibinfo  [0]{\@secondoftwo}%
\providecommand \bibfield  [0]{\@secondoftwo}%
\providecommand \translation [1]{[#1]}%
\providecommand \BibitemOpen [0]{}%
\providecommand \bibitemStop [0]{}%
\providecommand \bibitemNoStop [0]{.\EOS\space}%
\providecommand \EOS [0]{\spacefactor3000\relax}%
\providecommand \BibitemShut  [1]{\csname bibitem#1\endcsname}%
\let\auto@bib@innerbib\@empty
\bibitem [{\citenamefont {Carusotto}\ and\ \citenamefont
  {Ciuti}(2013)}]{Carusotto2013}%
  \BibitemOpen
  \bibfield  {author} {\bibinfo {author} {\bibfnamefont {I.}~\bibnamefont
  {Carusotto}}\ and\ \bibinfo {author} {\bibfnamefont {C.}~\bibnamefont
  {Ciuti}},\ }\href@noop {} {\bibfield  {journal} {\bibinfo  {journal} {Reviews
  of Modern Physics}\ }\textbf {\bibinfo {volume} {85}},\ \bibinfo {pages}
  {299} (\bibinfo {year} {2013})}\BibitemShut {NoStop}%
\bibitem [{\citenamefont {Sommer}\ \emph {et~al.}(2015)\citenamefont {Sommer},
  \citenamefont {B{\"u}chler},\ and\ \citenamefont {Simon}}]{Sommer2015}%
  \BibitemOpen
  \bibfield  {author} {\bibinfo {author} {\bibfnamefont {A.}~\bibnamefont
  {Sommer}}, \bibinfo {author} {\bibfnamefont {H.~P.}\ \bibnamefont
  {B{\"u}chler}}, \ and\ \bibinfo {author} {\bibfnamefont {J.}~\bibnamefont
  {Simon}},\ }\href@noop {} {\bibfield  {journal} {\bibinfo  {journal}
  {arXiv:1506.00341}\ } (\bibinfo {year} {2015})}\BibitemShut {NoStop}%
\bibitem [{\citenamefont {Peyronel}\ \emph {et~al.}(2012)\citenamefont
  {Peyronel}, \citenamefont {Firstenberg}, \citenamefont {Liang}, \citenamefont
  {Hofferberth}, \citenamefont {Gorshkov}, \citenamefont {Pohl}, \citenamefont
  {Lukin},\ and\ \citenamefont {Vuleti{\'c}}}]{Peyronel2012}%
  \BibitemOpen
  \bibfield  {author} {\bibinfo {author} {\bibfnamefont {T.}~\bibnamefont
  {Peyronel}}, \bibinfo {author} {\bibfnamefont {O.}~\bibnamefont
  {Firstenberg}}, \bibinfo {author} {\bibfnamefont {Q.-Y.}\ \bibnamefont
  {Liang}}, \bibinfo {author} {\bibfnamefont {S.}~\bibnamefont {Hofferberth}},
  \bibinfo {author} {\bibfnamefont {A.~V.}\ \bibnamefont {Gorshkov}}, \bibinfo
  {author} {\bibfnamefont {T.}~\bibnamefont {Pohl}}, \bibinfo {author}
  {\bibfnamefont {M.~D.}\ \bibnamefont {Lukin}}, \ and\ \bibinfo {author}
  {\bibfnamefont {V.}~\bibnamefont {Vuleti{\'c}}},\ }\href@noop {} {\bibfield
  {journal} {\bibinfo  {journal} {Nature}\ }\textbf {\bibinfo {volume} {488}},\
  \bibinfo {pages} {57} (\bibinfo {year} {2012})}\BibitemShut {NoStop}%
\bibitem [{\citenamefont {Weatherill}\ \emph {et~al.}(2008)\citenamefont
  {Weatherill}, \citenamefont {Pritchard}, \citenamefont {Abel}, \citenamefont
  {Bason}, \citenamefont {Mohapatra},\ and\ \citenamefont
  {Adams}}]{weatherill2008electromagnetically}%
  \BibitemOpen
  \bibfield  {author} {\bibinfo {author} {\bibfnamefont {K.}~\bibnamefont
  {Weatherill}}, \bibinfo {author} {\bibfnamefont {J.}~\bibnamefont
  {Pritchard}}, \bibinfo {author} {\bibfnamefont {R.}~\bibnamefont {Abel}},
  \bibinfo {author} {\bibfnamefont {M.}~\bibnamefont {Bason}}, \bibinfo
  {author} {\bibfnamefont {A.}~\bibnamefont {Mohapatra}}, \ and\ \bibinfo
  {author} {\bibfnamefont {C.}~\bibnamefont {Adams}},\ }\href@noop {}
  {\bibfield  {journal} {\bibinfo  {journal} {Journal of Physics B: Atomic,
  Molecular and Optical Physics}\ }\textbf {\bibinfo {volume} {41}},\ \bibinfo
  {pages} {201002} (\bibinfo {year} {2008})}\BibitemShut {NoStop}%
\bibitem [{\citenamefont {Petrosyan}\ \emph {et~al.}(2011)\citenamefont
  {Petrosyan}, \citenamefont {Otterbach},\ and\ \citenamefont
  {Fleischhauer}}]{Petrosyan2011}%
  \BibitemOpen
  \bibfield  {author} {\bibinfo {author} {\bibfnamefont {D.}~\bibnamefont
  {Petrosyan}}, \bibinfo {author} {\bibfnamefont {J.}~\bibnamefont
  {Otterbach}}, \ and\ \bibinfo {author} {\bibfnamefont {M.}~\bibnamefont
  {Fleischhauer}},\ }\href@noop {} {\bibfield  {journal} {\bibinfo  {journal}
  {Physical Review Letters}\ }\textbf {\bibinfo {volume} {107}},\ \bibinfo
  {pages} {213601} (\bibinfo {year} {2011})}\BibitemShut {NoStop}%
\bibitem [{\citenamefont {Schine}\ \emph {et~al.}(2016)\citenamefont {Schine},
  \citenamefont {Ryou}, \citenamefont {Gromov}, \citenamefont {Sommer},\ and\
  \citenamefont {Simon}}]{Schine2016}%
  \BibitemOpen
  \bibfield  {author} {\bibinfo {author} {\bibfnamefont {N.}~\bibnamefont
  {Schine}}, \bibinfo {author} {\bibfnamefont {A.}~\bibnamefont {Ryou}},
  \bibinfo {author} {\bibfnamefont {A.}~\bibnamefont {Gromov}}, \bibinfo
  {author} {\bibfnamefont {A.}~\bibnamefont {Sommer}}, \ and\ \bibinfo {author}
  {\bibfnamefont {J.}~\bibnamefont {Simon}},\ }\href@noop {} {\bibfield
  {journal} {\bibinfo  {journal} {Nature}\ }\textbf {\bibinfo {volume} {534}},\
  \bibinfo {pages} {671} (\bibinfo {year} {2016})}\BibitemShut {NoStop}%
\bibitem [{\citenamefont {Sommer}\ and\ \citenamefont
  {Simon}(2016)}]{Sommer2016}%
  \BibitemOpen
  \bibfield  {author} {\bibinfo {author} {\bibfnamefont {A.}~\bibnamefont
  {Sommer}}\ and\ \bibinfo {author} {\bibfnamefont {J.}~\bibnamefont {Simon}},\
  }\href@noop {} {\bibfield  {journal} {\bibinfo  {journal} {New Journal of
  Physics}\ }\textbf {\bibinfo {volume} {18}},\ \bibinfo {pages} {035008}
  (\bibinfo {year} {2016})}\BibitemShut {NoStop}%
\bibitem [{\citenamefont {Gorshkov}\ \emph {et~al.}(2011)\citenamefont
  {Gorshkov}, \citenamefont {Otterbach}, \citenamefont {Fleischhauer},
  \citenamefont {Pohl},\ and\ \citenamefont {Lukin}}]{Gorshkov2011}%
  \BibitemOpen
  \bibfield  {author} {\bibinfo {author} {\bibfnamefont {A.~V.}\ \bibnamefont
  {Gorshkov}}, \bibinfo {author} {\bibfnamefont {J.}~\bibnamefont {Otterbach}},
  \bibinfo {author} {\bibfnamefont {M.}~\bibnamefont {Fleischhauer}}, \bibinfo
  {author} {\bibfnamefont {T.}~\bibnamefont {Pohl}}, \ and\ \bibinfo {author}
  {\bibfnamefont {M.~D.}\ \bibnamefont {Lukin}},\ }\href@noop {} {\bibfield
  {journal} {\bibinfo  {journal} {Physical Review Letters}\ }\textbf {\bibinfo
  {volume} {107}},\ \bibinfo {pages} {133602} (\bibinfo {year}
  {2011})}\BibitemShut {NoStop}%
\bibitem [{\citenamefont {Gullans}\ \emph {et~al.}(2016)\citenamefont
  {Gullans}, \citenamefont {Thompson}, \citenamefont {Wang}, \citenamefont
  {Liang}, \citenamefont {Vuleti{\'c}}, \citenamefont {Lukin},\ and\
  \citenamefont {Gorshkov}}]{Gullans2016}%
  \BibitemOpen
  \bibfield  {author} {\bibinfo {author} {\bibfnamefont {M.}~\bibnamefont
  {Gullans}}, \bibinfo {author} {\bibfnamefont {J.}~\bibnamefont {Thompson}},
  \bibinfo {author} {\bibfnamefont {Y.}~\bibnamefont {Wang}}, \bibinfo {author}
  {\bibfnamefont {Q.-Y.}\ \bibnamefont {Liang}}, \bibinfo {author}
  {\bibfnamefont {V.}~\bibnamefont {Vuleti{\'c}}}, \bibinfo {author}
  {\bibfnamefont {M.~D.}\ \bibnamefont {Lukin}}, \ and\ \bibinfo {author}
  {\bibfnamefont {A.~V.}\ \bibnamefont {Gorshkov}},\ }\href@noop {} {\bibfield
  {journal} {\bibinfo  {journal} {Physical Review Letters}\ }\textbf {\bibinfo
  {volume} {117}},\ \bibinfo {pages} {113601} (\bibinfo {year}
  {2016})}\BibitemShut {NoStop}%
\bibitem [{\citenamefont {Umucal{\i}lar}\ \emph {et~al.}(2014)\citenamefont
  {Umucal{\i}lar}, \citenamefont {Wouters},\ and\ \citenamefont
  {Carusotto}}]{Umucalilar2014}%
  \BibitemOpen
  \bibfield  {author} {\bibinfo {author} {\bibfnamefont {R.}~\bibnamefont
  {Umucal{\i}lar}}, \bibinfo {author} {\bibfnamefont {M.}~\bibnamefont
  {Wouters}}, \ and\ \bibinfo {author} {\bibfnamefont {I.}~\bibnamefont
  {Carusotto}},\ }\href@noop {} {\bibfield  {journal} {\bibinfo  {journal}
  {Physical Review A}\ }\textbf {\bibinfo {volume} {89}},\ \bibinfo {pages}
  {023803} (\bibinfo {year} {2014})}\BibitemShut {NoStop}%
\bibitem [{\citenamefont {Grusdt}\ and\ \citenamefont
  {Fleischhauer}(2013)}]{Grusdt2013}%
  \BibitemOpen
  \bibfield  {author} {\bibinfo {author} {\bibfnamefont {F.}~\bibnamefont
  {Grusdt}}\ and\ \bibinfo {author} {\bibfnamefont {M.}~\bibnamefont
  {Fleischhauer}},\ }\href@noop {} {\bibfield  {journal} {\bibinfo  {journal}
  {Physical Review A}\ }\textbf {\bibinfo {volume} {87}},\ \bibinfo {pages}
  {043628} (\bibinfo {year} {2013})}\BibitemShut {NoStop}%
\bibitem [{\citenamefont {Jaksch}\ \emph {et~al.}(2000)\citenamefont {Jaksch},
  \citenamefont {Cirac}, \citenamefont {Zoller}, \citenamefont {Rolston},
  \citenamefont {C{\^o}t{\'e}},\ and\ \citenamefont {Lukin}}]{Jaksch2000}%
  \BibitemOpen
  \bibfield  {author} {\bibinfo {author} {\bibfnamefont {D.}~\bibnamefont
  {Jaksch}}, \bibinfo {author} {\bibfnamefont {J.}~\bibnamefont {Cirac}},
  \bibinfo {author} {\bibfnamefont {P.}~\bibnamefont {Zoller}}, \bibinfo
  {author} {\bibfnamefont {S.}~\bibnamefont {Rolston}}, \bibinfo {author}
  {\bibfnamefont {R.}~\bibnamefont {C{\^o}t{\'e}}}, \ and\ \bibinfo {author}
  {\bibfnamefont {M.}~\bibnamefont {Lukin}},\ }\href@noop {} {\bibfield
  {journal} {\bibinfo  {journal} {Physical Review Letters}\ }\textbf {\bibinfo
  {volume} {85}},\ \bibinfo {pages} {2208} (\bibinfo {year}
  {2000})}\BibitemShut {NoStop}%
\bibitem [{\citenamefont {Saffman}\ \emph {et~al.}(2010)\citenamefont
  {Saffman}, \citenamefont {Walker},\ and\ \citenamefont
  {M{\o}lmer}}]{Saffman2010}%
  \BibitemOpen
  \bibfield  {author} {\bibinfo {author} {\bibfnamefont {M.}~\bibnamefont
  {Saffman}}, \bibinfo {author} {\bibfnamefont {T.~G.}\ \bibnamefont {Walker}},
  \ and\ \bibinfo {author} {\bibfnamefont {K.}~\bibnamefont {M{\o}lmer}},\
  }\href@noop {} {\bibfield  {journal} {\bibinfo  {journal} {Reviews of Modern
  Physics}\ }\textbf {\bibinfo {volume} {82}},\ \bibinfo {pages} {2313}
  (\bibinfo {year} {2010})}\BibitemShut {NoStop}%
\bibitem [{\citenamefont {Han}\ \emph {et~al.}(2010)\citenamefont {Han},
  \citenamefont {He}, \citenamefont {Heshami}, \citenamefont {Li},\ and\
  \citenamefont {Simon}}]{han2010quantum}%
  \BibitemOpen
  \bibfield  {author} {\bibinfo {author} {\bibfnamefont {Y.}~\bibnamefont
  {Han}}, \bibinfo {author} {\bibfnamefont {B.}~\bibnamefont {He}}, \bibinfo
  {author} {\bibfnamefont {K.}~\bibnamefont {Heshami}}, \bibinfo {author}
  {\bibfnamefont {C.-Z.}\ \bibnamefont {Li}}, \ and\ \bibinfo {author}
  {\bibfnamefont {C.}~\bibnamefont {Simon}},\ }\href@noop {} {\bibfield
  {journal} {\bibinfo  {journal} {Physical Review A}\ }\textbf {\bibinfo
  {volume} {81}},\ \bibinfo {pages} {052311} (\bibinfo {year}
  {2010})}\BibitemShut {NoStop}%
\bibitem [{\citenamefont {Stormer}\ \emph {et~al.}(1999)\citenamefont
  {Stormer}, \citenamefont {Tsui},\ and\ \citenamefont
  {Gossard}}]{Stormer1999}%
  \BibitemOpen
  \bibfield  {author} {\bibinfo {author} {\bibfnamefont {H.~L.}\ \bibnamefont
  {Stormer}}, \bibinfo {author} {\bibfnamefont {D.~C.}\ \bibnamefont {Tsui}}, \
  and\ \bibinfo {author} {\bibfnamefont {A.~C.}\ \bibnamefont {Gossard}},\
  }\href@noop {} {\bibfield  {journal} {\bibinfo  {journal} {Reviews of Modern
  Physics}\ }\textbf {\bibinfo {volume} {71}},\ \bibinfo {pages} {S298}
  (\bibinfo {year} {1999})}\BibitemShut {NoStop}%
\bibitem [{\citenamefont {Houck}\ \emph {et~al.}(2012)\citenamefont {Houck},
  \citenamefont {T{\"u}reci},\ and\ \citenamefont {Koch}}]{Houck2012}%
  \BibitemOpen
  \bibfield  {author} {\bibinfo {author} {\bibfnamefont {A.~A.}\ \bibnamefont
  {Houck}}, \bibinfo {author} {\bibfnamefont {H.~E.}\ \bibnamefont
  {T{\"u}reci}}, \ and\ \bibinfo {author} {\bibfnamefont {J.}~\bibnamefont
  {Koch}},\ }\href@noop {} {\bibfield  {journal} {\bibinfo  {journal} {Nature
  Physics}\ }\textbf {\bibinfo {volume} {8}},\ \bibinfo {pages} {292} (\bibinfo
  {year} {2012})}\BibitemShut {NoStop}%
\bibitem [{\citenamefont {Anderson}\ \emph {et~al.}(2016)\citenamefont
  {Anderson}, \citenamefont {Ma}, \citenamefont {Owens}, \citenamefont
  {Schuster},\ and\ \citenamefont {Simon}}]{Anderson2016}%
  \BibitemOpen
  \bibfield  {author} {\bibinfo {author} {\bibfnamefont {B.~M.}\ \bibnamefont
  {Anderson}}, \bibinfo {author} {\bibfnamefont {R.}~\bibnamefont {Ma}},
  \bibinfo {author} {\bibfnamefont {C.}~\bibnamefont {Owens}}, \bibinfo
  {author} {\bibfnamefont {D.~I.}\ \bibnamefont {Schuster}}, \ and\ \bibinfo
  {author} {\bibfnamefont {J.}~\bibnamefont {Simon}},\ }\href@noop {}
  {\bibfield  {journal} {\bibinfo  {journal} {Physical Review X}\ }\textbf
  {\bibinfo {volume} {6}},\ \bibinfo {pages} {041043} (\bibinfo {year}
  {2016})}\BibitemShut {NoStop}%
\bibitem [{\citenamefont {Roushan}\ \emph {et~al.}(2017)\citenamefont
  {Roushan}, \citenamefont {Neill}, \citenamefont {Megrant}, \citenamefont
  {Chen}, \citenamefont {Babbush}, \citenamefont {Barends}, \citenamefont
  {Campbell}, \citenamefont {Chen}, \citenamefont {Chiaro}, \citenamefont
  {Dunsworth} \emph {et~al.}}]{Roushan2017}%
  \BibitemOpen
  \bibfield  {author} {\bibinfo {author} {\bibfnamefont {P.}~\bibnamefont
  {Roushan}}, \bibinfo {author} {\bibfnamefont {C.}~\bibnamefont {Neill}},
  \bibinfo {author} {\bibfnamefont {A.}~\bibnamefont {Megrant}}, \bibinfo
  {author} {\bibfnamefont {Y.}~\bibnamefont {Chen}}, \bibinfo {author}
  {\bibfnamefont {R.}~\bibnamefont {Babbush}}, \bibinfo {author} {\bibfnamefont
  {R.}~\bibnamefont {Barends}}, \bibinfo {author} {\bibfnamefont
  {B.}~\bibnamefont {Campbell}}, \bibinfo {author} {\bibfnamefont
  {Z.}~\bibnamefont {Chen}}, \bibinfo {author} {\bibfnamefont {B.}~\bibnamefont
  {Chiaro}}, \bibinfo {author} {\bibfnamefont {A.}~\bibnamefont {Dunsworth}},
  \emph {et~al.},\ }\href@noop {} {\bibfield  {journal} {\bibinfo  {journal}
  {Nature Physics}\ }\textbf {\bibinfo {volume} {13}},\ \bibinfo {pages} {146}
  (\bibinfo {year} {2017})}\BibitemShut {NoStop}%
\bibitem [{\citenamefont {Tsui}\ \emph {et~al.}(1982)\citenamefont {Tsui},
  \citenamefont {Stormer},\ and\ \citenamefont {Gossard}}]{Tsui1982}%
  \BibitemOpen
  \bibfield  {author} {\bibinfo {author} {\bibfnamefont {D.~C.}\ \bibnamefont
  {Tsui}}, \bibinfo {author} {\bibfnamefont {H.~L.}\ \bibnamefont {Stormer}}, \
  and\ \bibinfo {author} {\bibfnamefont {A.~C.}\ \bibnamefont {Gossard}},\
  }\href@noop {} {\bibfield  {journal} {\bibinfo  {journal} {Physical Review
  Letters}\ }\textbf {\bibinfo {volume} {48}},\ \bibinfo {pages} {1559}
  (\bibinfo {year} {1982})}\BibitemShut {NoStop}%
\bibitem [{\citenamefont {Deng}\ \emph {et~al.}(2010)\citenamefont {Deng},
  \citenamefont {Haug},\ and\ \citenamefont {Yamamoto}}]{Deng2010}%
  \BibitemOpen
  \bibfield  {author} {\bibinfo {author} {\bibfnamefont {H.}~\bibnamefont
  {Deng}}, \bibinfo {author} {\bibfnamefont {H.}~\bibnamefont {Haug}}, \ and\
  \bibinfo {author} {\bibfnamefont {Y.}~\bibnamefont {Yamamoto}},\ }\href@noop
  {} {\bibfield  {journal} {\bibinfo  {journal} {Reviews of Modern Physics}\
  }\textbf {\bibinfo {volume} {82}},\ \bibinfo {pages} {1489} (\bibinfo {year}
  {2010})}\BibitemShut {NoStop}%
\bibitem [{\citenamefont {Dean}\ \emph {et~al.}(2010)\citenamefont {Dean},
  \citenamefont {Young}, \citenamefont {Meric}, \citenamefont {Lee},
  \citenamefont {Wang}, \citenamefont {Sorgenfrei}, \citenamefont {Watanabe},
  \citenamefont {Taniguchi}, \citenamefont {Kim}, \citenamefont {Shepard} \emph
  {et~al.}}]{Dean2010}%
  \BibitemOpen
  \bibfield  {author} {\bibinfo {author} {\bibfnamefont {C.~R.}\ \bibnamefont
  {Dean}}, \bibinfo {author} {\bibfnamefont {A.~F.}\ \bibnamefont {Young}},
  \bibinfo {author} {\bibfnamefont {I.}~\bibnamefont {Meric}}, \bibinfo
  {author} {\bibfnamefont {C.}~\bibnamefont {Lee}}, \bibinfo {author}
  {\bibfnamefont {L.}~\bibnamefont {Wang}}, \bibinfo {author} {\bibfnamefont
  {S.}~\bibnamefont {Sorgenfrei}}, \bibinfo {author} {\bibfnamefont
  {K.}~\bibnamefont {Watanabe}}, \bibinfo {author} {\bibfnamefont
  {T.}~\bibnamefont {Taniguchi}}, \bibinfo {author} {\bibfnamefont
  {P.}~\bibnamefont {Kim}}, \bibinfo {author} {\bibfnamefont {K.~L.}\
  \bibnamefont {Shepard}},  \emph {et~al.},\ }\href@noop {} {\bibfield
  {journal} {\bibinfo  {journal} {Nature Nanotechnology}\ }\textbf {\bibinfo
  {volume} {5}},\ \bibinfo {pages} {722} (\bibinfo {year} {2010})}\BibitemShut
  {NoStop}%
\bibitem [{\citenamefont {Greiner}\ \emph {et~al.}(2002)\citenamefont
  {Greiner}, \citenamefont {Mandel}, \citenamefont {Esslinger}, \citenamefont
  {H{\"a}nsch},\ and\ \citenamefont {Bloch}}]{Greiner2002}%
  \BibitemOpen
  \bibfield  {author} {\bibinfo {author} {\bibfnamefont {M.}~\bibnamefont
  {Greiner}}, \bibinfo {author} {\bibfnamefont {O.}~\bibnamefont {Mandel}},
  \bibinfo {author} {\bibfnamefont {T.}~\bibnamefont {Esslinger}}, \bibinfo
  {author} {\bibfnamefont {T.~W.}\ \bibnamefont {H{\"a}nsch}}, \ and\ \bibinfo
  {author} {\bibfnamefont {I.}~\bibnamefont {Bloch}},\ }\href@noop {}
  {\bibfield  {journal} {\bibinfo  {journal} {Nature}\ }\textbf {\bibinfo
  {volume} {415}},\ \bibinfo {pages} {39} (\bibinfo {year} {2002})}\BibitemShut
  {NoStop}%
\bibitem [{\citenamefont {Inouye}\ \emph {et~al.}(1998)\citenamefont {Inouye},
  \citenamefont {Andrews}, \citenamefont {Stenger}, \citenamefont {Miesner},
  \citenamefont {Stamper-Kurn},\ and\ \citenamefont {Ketterle}}]{Inouye1998}%
  \BibitemOpen
  \bibfield  {author} {\bibinfo {author} {\bibfnamefont {S.}~\bibnamefont
  {Inouye}}, \bibinfo {author} {\bibfnamefont {M.}~\bibnamefont {Andrews}},
  \bibinfo {author} {\bibfnamefont {J.}~\bibnamefont {Stenger}}, \bibinfo
  {author} {\bibfnamefont {H.-J.}\ \bibnamefont {Miesner}}, \bibinfo {author}
  {\bibfnamefont {D.}~\bibnamefont {Stamper-Kurn}}, \ and\ \bibinfo {author}
  {\bibfnamefont {W.}~\bibnamefont {Ketterle}},\ }\href@noop {} {\bibfield
  {journal} {\bibinfo  {journal} {Nature}\ }\textbf {\bibinfo {volume} {392}},\
  \bibinfo {pages} {151} (\bibinfo {year} {1998})}\BibitemShut {NoStop}%
\bibitem [{\citenamefont {Vuleti{\'c}}\ \emph {et~al.}(1999)\citenamefont
  {Vuleti{\'c}}, \citenamefont {Kerman}, \citenamefont {Chin},\ and\
  \citenamefont {Chu}}]{Vuletic1999}%
  \BibitemOpen
  \bibfield  {author} {\bibinfo {author} {\bibfnamefont {V.}~\bibnamefont
  {Vuleti{\'c}}}, \bibinfo {author} {\bibfnamefont {A.~J.}\ \bibnamefont
  {Kerman}}, \bibinfo {author} {\bibfnamefont {C.}~\bibnamefont {Chin}}, \ and\
  \bibinfo {author} {\bibfnamefont {S.}~\bibnamefont {Chu}},\ }\href@noop {}
  {\bibfield  {journal} {\bibinfo  {journal} {Physical Review Letters}\
  }\textbf {\bibinfo {volume} {82}},\ \bibinfo {pages} {1406} (\bibinfo {year}
  {1999})}\BibitemShut {NoStop}%
\bibitem [{\citenamefont {Schreier}\ \emph {et~al.}(2008)\citenamefont
  {Schreier}, \citenamefont {Houck}, \citenamefont {Koch}, \citenamefont
  {Schuster}, \citenamefont {Johnson}, \citenamefont {Chow}, \citenamefont
  {Gambetta}, \citenamefont {Majer}, \citenamefont {Frunzio}, \citenamefont
  {Devoret} \emph {et~al.}}]{Schreier2008}%
  \BibitemOpen
  \bibfield  {author} {\bibinfo {author} {\bibfnamefont {J.}~\bibnamefont
  {Schreier}}, \bibinfo {author} {\bibfnamefont {A.~A.}\ \bibnamefont {Houck}},
  \bibinfo {author} {\bibfnamefont {J.}~\bibnamefont {Koch}}, \bibinfo {author}
  {\bibfnamefont {D.~I.}\ \bibnamefont {Schuster}}, \bibinfo {author}
  {\bibfnamefont {B.}~\bibnamefont {Johnson}}, \bibinfo {author} {\bibfnamefont
  {J.}~\bibnamefont {Chow}}, \bibinfo {author} {\bibfnamefont {J.~M.}\
  \bibnamefont {Gambetta}}, \bibinfo {author} {\bibfnamefont {J.}~\bibnamefont
  {Majer}}, \bibinfo {author} {\bibfnamefont {L.}~\bibnamefont {Frunzio}},
  \bibinfo {author} {\bibfnamefont {M.~H.}\ \bibnamefont {Devoret}},  \emph
  {et~al.},\ }\href@noop {} {\bibfield  {journal} {\bibinfo  {journal}
  {Physical Review B}\ }\textbf {\bibinfo {volume} {77}},\ \bibinfo {pages}
  {180502} (\bibinfo {year} {2008})}\BibitemShut {NoStop}%
\bibitem [{\citenamefont {Schuster}\ \emph {et~al.}(2007)\citenamefont
  {Schuster}, \citenamefont {Houck}, \citenamefont {Schreier}, \citenamefont
  {Wallraff}, \citenamefont {Gambetta}, \citenamefont {Blais}, \citenamefont
  {Frunzio}, \citenamefont {Majer}, \citenamefont {Johnson}, \citenamefont
  {Devoret} \emph {et~al.}}]{Schuster2007}%
  \BibitemOpen
  \bibfield  {author} {\bibinfo {author} {\bibfnamefont {D.}~\bibnamefont
  {Schuster}}, \bibinfo {author} {\bibfnamefont {A.}~\bibnamefont {Houck}},
  \bibinfo {author} {\bibfnamefont {J.}~\bibnamefont {Schreier}}, \bibinfo
  {author} {\bibfnamefont {A.}~\bibnamefont {Wallraff}}, \bibinfo {author}
  {\bibfnamefont {J.}~\bibnamefont {Gambetta}}, \bibinfo {author}
  {\bibfnamefont {A.}~\bibnamefont {Blais}}, \bibinfo {author} {\bibfnamefont
  {L.}~\bibnamefont {Frunzio}}, \bibinfo {author} {\bibfnamefont
  {J.}~\bibnamefont {Majer}}, \bibinfo {author} {\bibfnamefont
  {B.}~\bibnamefont {Johnson}}, \bibinfo {author} {\bibfnamefont
  {M.}~\bibnamefont {Devoret}},  \emph {et~al.},\ }\href@noop {} {\bibfield
  {journal} {\bibinfo  {journal} {Nature}\ }\textbf {\bibinfo {volume} {445}},\
  \bibinfo {pages} {515} (\bibinfo {year} {2007})}\BibitemShut {NoStop}%
\bibitem [{\citenamefont {Underwood}\ \emph {et~al.}(2012)\citenamefont
  {Underwood}, \citenamefont {Shanks}, \citenamefont {Koch},\ and\
  \citenamefont {Houck}}]{Underwood2012}%
  \BibitemOpen
  \bibfield  {author} {\bibinfo {author} {\bibfnamefont {D.~L.}\ \bibnamefont
  {Underwood}}, \bibinfo {author} {\bibfnamefont {W.~E.}\ \bibnamefont
  {Shanks}}, \bibinfo {author} {\bibfnamefont {J.}~\bibnamefont {Koch}}, \ and\
  \bibinfo {author} {\bibfnamefont {A.~A.}\ \bibnamefont {Houck}},\ }\href@noop
  {} {\bibfield  {journal} {\bibinfo  {journal} {Physical Review A}\ }\textbf
  {\bibinfo {volume} {86}},\ \bibinfo {pages} {023837} (\bibinfo {year}
  {2012})}\BibitemShut {NoStop}%
\bibitem [{\citenamefont {Owens}\ \emph {et~al.}(2018)\citenamefont {Owens},
  \citenamefont {LaChapelle}, \citenamefont {Saxberg}, \citenamefont
  {Anderson}, \citenamefont {Ma}, \citenamefont {Simon},\ and\ \citenamefont
  {Schuster}}]{Owens2018}%
  \BibitemOpen
  \bibfield  {author} {\bibinfo {author} {\bibfnamefont {C.}~\bibnamefont
  {Owens}}, \bibinfo {author} {\bibfnamefont {A.}~\bibnamefont {LaChapelle}},
  \bibinfo {author} {\bibfnamefont {B.}~\bibnamefont {Saxberg}}, \bibinfo
  {author} {\bibfnamefont {B.~M.}\ \bibnamefont {Anderson}}, \bibinfo {author}
  {\bibfnamefont {R.}~\bibnamefont {Ma}}, \bibinfo {author} {\bibfnamefont
  {J.}~\bibnamefont {Simon}}, \ and\ \bibinfo {author} {\bibfnamefont {D.~I.}\
  \bibnamefont {Schuster}},\ }\href@noop {} {\bibfield  {journal} {\bibinfo
  {journal} {Physical Review A}\ }\textbf {\bibinfo {volume} {97}},\ \bibinfo
  {pages} {013818} (\bibinfo {year} {2018})}\BibitemShut {NoStop}%
\bibitem [{\citenamefont {Fitzpatrick}\ \emph {et~al.}(2017)\citenamefont
  {Fitzpatrick}, \citenamefont {Sundaresan}, \citenamefont {Li}, \citenamefont
  {Koch},\ and\ \citenamefont {Houck}}]{Fitzpatrick2017}%
  \BibitemOpen
  \bibfield  {author} {\bibinfo {author} {\bibfnamefont {M.}~\bibnamefont
  {Fitzpatrick}}, \bibinfo {author} {\bibfnamefont {N.~M.}\ \bibnamefont
  {Sundaresan}}, \bibinfo {author} {\bibfnamefont {A.~C.}\ \bibnamefont {Li}},
  \bibinfo {author} {\bibfnamefont {J.}~\bibnamefont {Koch}}, \ and\ \bibinfo
  {author} {\bibfnamefont {A.~A.}\ \bibnamefont {Houck}},\ }\href@noop {}
  {\bibfield  {journal} {\bibinfo  {journal} {Physical Review X}\ }\textbf
  {\bibinfo {volume} {7}},\ \bibinfo {pages} {011016} (\bibinfo {year}
  {2017})}\BibitemShut {NoStop}%
\bibitem [{\citenamefont {Klaers}\ \emph {et~al.}(2010)\citenamefont {Klaers},
  \citenamefont {Schmitt}, \citenamefont {Vewinger},\ and\ \citenamefont
  {Weitz}}]{Klaers2010}%
  \BibitemOpen
  \bibfield  {author} {\bibinfo {author} {\bibfnamefont {J.}~\bibnamefont
  {Klaers}}, \bibinfo {author} {\bibfnamefont {J.}~\bibnamefont {Schmitt}},
  \bibinfo {author} {\bibfnamefont {F.}~\bibnamefont {Vewinger}}, \ and\
  \bibinfo {author} {\bibfnamefont {M.}~\bibnamefont {Weitz}},\ }\href@noop {}
  {\bibfield  {journal} {\bibinfo  {journal} {Nature}\ }\textbf {\bibinfo
  {volume} {468}},\ \bibinfo {pages} {545} (\bibinfo {year}
  {2010})}\BibitemShut {NoStop}%
\bibitem [{\citenamefont {Carroll}\ \emph {et~al.}(2004)\citenamefont
  {Carroll}, \citenamefont {Claringbould}, \citenamefont {Goodsell},
  \citenamefont {Lim},\ and\ \citenamefont {Noel}}]{Carroll2004}%
  \BibitemOpen
  \bibfield  {author} {\bibinfo {author} {\bibfnamefont {T.~J.}\ \bibnamefont
  {Carroll}}, \bibinfo {author} {\bibfnamefont {K.}~\bibnamefont
  {Claringbould}}, \bibinfo {author} {\bibfnamefont {A.}~\bibnamefont
  {Goodsell}}, \bibinfo {author} {\bibfnamefont {M.}~\bibnamefont {Lim}}, \
  and\ \bibinfo {author} {\bibfnamefont {M.~W.}\ \bibnamefont {Noel}},\
  }\href@noop {} {\bibfield  {journal} {\bibinfo  {journal} {Physical Review
  Letters}\ }\textbf {\bibinfo {volume} {93}},\ \bibinfo {pages} {153001}
  (\bibinfo {year} {2004})}\BibitemShut {NoStop}%
\bibitem [{\citenamefont {Firstenberg}\ \emph {et~al.}(2013)\citenamefont
  {Firstenberg}, \citenamefont {Peyronel}, \citenamefont {Liang}, \citenamefont
  {Gorshkov}, \citenamefont {Lukin},\ and\ \citenamefont
  {Vuleti{\'c}}}]{Firstenberg2013}%
  \BibitemOpen
  \bibfield  {author} {\bibinfo {author} {\bibfnamefont {O.}~\bibnamefont
  {Firstenberg}}, \bibinfo {author} {\bibfnamefont {T.}~\bibnamefont
  {Peyronel}}, \bibinfo {author} {\bibfnamefont {Q.-Y.}\ \bibnamefont {Liang}},
  \bibinfo {author} {\bibfnamefont {A.~V.}\ \bibnamefont {Gorshkov}}, \bibinfo
  {author} {\bibfnamefont {M.~D.}\ \bibnamefont {Lukin}}, \ and\ \bibinfo
  {author} {\bibfnamefont {V.}~\bibnamefont {Vuleti{\'c}}},\ }\href@noop {}
  {\bibfield  {journal} {\bibinfo  {journal} {Nature}\ }\textbf {\bibinfo
  {volume} {502}},\ \bibinfo {pages} {71} (\bibinfo {year} {2013})}\BibitemShut
  {NoStop}%
\bibitem [{\citenamefont {Parigi}\ \emph {et~al.}(2012)\citenamefont {Parigi},
  \citenamefont {Bimbard}, \citenamefont {Stanojevic}, \citenamefont
  {Hilliard}, \citenamefont {Nogrette}, \citenamefont {Tualle-Brouri},
  \citenamefont {Ourjoumtsev},\ and\ \citenamefont {Grangier}}]{Parigi2012}%
  \BibitemOpen
  \bibfield  {author} {\bibinfo {author} {\bibfnamefont {V.}~\bibnamefont
  {Parigi}}, \bibinfo {author} {\bibfnamefont {E.}~\bibnamefont {Bimbard}},
  \bibinfo {author} {\bibfnamefont {J.}~\bibnamefont {Stanojevic}}, \bibinfo
  {author} {\bibfnamefont {A.~J.}\ \bibnamefont {Hilliard}}, \bibinfo {author}
  {\bibfnamefont {F.}~\bibnamefont {Nogrette}}, \bibinfo {author}
  {\bibfnamefont {R.}~\bibnamefont {Tualle-Brouri}}, \bibinfo {author}
  {\bibfnamefont {A.}~\bibnamefont {Ourjoumtsev}}, \ and\ \bibinfo {author}
  {\bibfnamefont {P.}~\bibnamefont {Grangier}},\ }\href@noop {} {\bibfield
  {journal} {\bibinfo  {journal} {Physical Review Letters}\ }\textbf {\bibinfo
  {volume} {109}},\ \bibinfo {pages} {233602} (\bibinfo {year}
  {2012})}\BibitemShut {NoStop}%
\bibitem [{\citenamefont {Stanojevic}\ \emph {et~al.}(2013)\citenamefont
  {Stanojevic}, \citenamefont {Parigi}, \citenamefont {Bimbard}, \citenamefont
  {Ourjoumtsev},\ and\ \citenamefont {Grangier}}]{stanojevic2013dispersive}%
  \BibitemOpen
  \bibfield  {author} {\bibinfo {author} {\bibfnamefont {J.}~\bibnamefont
  {Stanojevic}}, \bibinfo {author} {\bibfnamefont {V.}~\bibnamefont {Parigi}},
  \bibinfo {author} {\bibfnamefont {E.}~\bibnamefont {Bimbard}}, \bibinfo
  {author} {\bibfnamefont {A.}~\bibnamefont {Ourjoumtsev}}, \ and\ \bibinfo
  {author} {\bibfnamefont {P.}~\bibnamefont {Grangier}},\ }\href@noop {}
  {\bibfield  {journal} {\bibinfo  {journal} {Physical Review A}\ }\textbf
  {\bibinfo {volume} {88}},\ \bibinfo {pages} {053845} (\bibinfo {year}
  {2013})}\BibitemShut {NoStop}%
\bibitem [{\citenamefont {Ningyuan}\ \emph {et~al.}(2016)\citenamefont
  {Ningyuan}, \citenamefont {Georgakopoulos}, \citenamefont {Ryou},
  \citenamefont {Schine}, \citenamefont {Sommer},\ and\ \citenamefont
  {Simon}}]{Jia2016}%
  \BibitemOpen
  \bibfield  {author} {\bibinfo {author} {\bibfnamefont {J.}~\bibnamefont
  {Ningyuan}}, \bibinfo {author} {\bibfnamefont {A.}~\bibnamefont
  {Georgakopoulos}}, \bibinfo {author} {\bibfnamefont {A.}~\bibnamefont
  {Ryou}}, \bibinfo {author} {\bibfnamefont {N.}~\bibnamefont {Schine}},
  \bibinfo {author} {\bibfnamefont {A.}~\bibnamefont {Sommer}}, \ and\ \bibinfo
  {author} {\bibfnamefont {J.}~\bibnamefont {Simon}},\ }\href@noop {}
  {\bibfield  {journal} {\bibinfo  {journal} {Physical Review A}\ }\textbf
  {\bibinfo {volume} {93}},\ \bibinfo {pages} {041802} (\bibinfo {year}
  {2016})}\BibitemShut {NoStop}%
\bibitem [{\citenamefont {Jia}\ \emph {et~al.}(2018)\citenamefont {Jia},
  \citenamefont {Schine}, \citenamefont {Georgakopoulos}, \citenamefont {Ryou},
  \citenamefont {Clark}, \citenamefont {Sommer},\ and\ \citenamefont
  {Simon}}]{Jia2018}%
  \BibitemOpen
  \bibfield  {author} {\bibinfo {author} {\bibfnamefont {N.}~\bibnamefont
  {Jia}}, \bibinfo {author} {\bibfnamefont {N.}~\bibnamefont {Schine}},
  \bibinfo {author} {\bibfnamefont {A.}~\bibnamefont {Georgakopoulos}},
  \bibinfo {author} {\bibfnamefont {A.}~\bibnamefont {Ryou}}, \bibinfo {author}
  {\bibfnamefont {L.~W.}\ \bibnamefont {Clark}}, \bibinfo {author}
  {\bibfnamefont {A.}~\bibnamefont {Sommer}}, \ and\ \bibinfo {author}
  {\bibfnamefont {J.}~\bibnamefont {Simon}},\ }\href@noop {} {\bibfield
  {journal} {\bibinfo  {journal} {Nature Physics}\ } (\bibinfo {year}
  {2018})}\BibitemShut {NoStop}%
\bibitem [{\citenamefont {Gorshkov}\ \emph {et~al.}(2007)\citenamefont
  {Gorshkov}, \citenamefont {Andr{\'e}}, \citenamefont {Lukin},\ and\
  \citenamefont {S{\o}rensen}}]{Gorshkov2007c}%
  \BibitemOpen
  \bibfield  {author} {\bibinfo {author} {\bibfnamefont {A.~V.}\ \bibnamefont
  {Gorshkov}}, \bibinfo {author} {\bibfnamefont {A.}~\bibnamefont {Andr{\'e}}},
  \bibinfo {author} {\bibfnamefont {M.~D.}\ \bibnamefont {Lukin}}, \ and\
  \bibinfo {author} {\bibfnamefont {A.~S.}\ \bibnamefont {S{\o}rensen}},\
  }\href@noop {} {\bibfield  {journal} {\bibinfo  {journal} {Physical Review
  A}\ }\textbf {\bibinfo {volume} {76}},\ \bibinfo {pages} {033805} (\bibinfo
  {year} {2007})}\BibitemShut {NoStop}%
\bibitem [{\citenamefont {Bienias}\ \emph {et~al.}(2014)\citenamefont
  {Bienias}, \citenamefont {Choi}, \citenamefont {Firstenberg}, \citenamefont
  {Maghrebi}, \citenamefont {Gullans}, \citenamefont {Lukin}, \citenamefont
  {Gorshkov},\ and\ \citenamefont {B{\"u}chler}}]{Bienias2014}%
  \BibitemOpen
  \bibfield  {author} {\bibinfo {author} {\bibfnamefont {P.}~\bibnamefont
  {Bienias}}, \bibinfo {author} {\bibfnamefont {S.}~\bibnamefont {Choi}},
  \bibinfo {author} {\bibfnamefont {O.}~\bibnamefont {Firstenberg}}, \bibinfo
  {author} {\bibfnamefont {M.}~\bibnamefont {Maghrebi}}, \bibinfo {author}
  {\bibfnamefont {M.}~\bibnamefont {Gullans}}, \bibinfo {author} {\bibfnamefont
  {M.~D.}\ \bibnamefont {Lukin}}, \bibinfo {author} {\bibfnamefont {A.~V.}\
  \bibnamefont {Gorshkov}}, \ and\ \bibinfo {author} {\bibfnamefont
  {H.}~\bibnamefont {B{\"u}chler}},\ }\href@noop {} {\bibfield  {journal}
  {\bibinfo  {journal} {Physical Review A}\ }\textbf {\bibinfo {volume} {90}},\
  \bibinfo {pages} {053804} (\bibinfo {year} {2014})}\BibitemShut {NoStop}%
\bibitem [{\citenamefont {Lukin}\ \emph {et~al.}(2001)\citenamefont {Lukin},
  \citenamefont {Fleischhauer}, \citenamefont {Cote}, \citenamefont {Duan},
  \citenamefont {Jaksch}, \citenamefont {Cirac},\ and\ \citenamefont
  {Zoller}}]{Lukin2001}%
  \BibitemOpen
  \bibfield  {author} {\bibinfo {author} {\bibfnamefont {M.}~\bibnamefont
  {Lukin}}, \bibinfo {author} {\bibfnamefont {M.}~\bibnamefont {Fleischhauer}},
  \bibinfo {author} {\bibfnamefont {R.}~\bibnamefont {Cote}}, \bibinfo {author}
  {\bibfnamefont {L.}~\bibnamefont {Duan}}, \bibinfo {author} {\bibfnamefont
  {D.}~\bibnamefont {Jaksch}}, \bibinfo {author} {\bibfnamefont
  {J.}~\bibnamefont {Cirac}}, \ and\ \bibinfo {author} {\bibfnamefont
  {P.}~\bibnamefont {Zoller}},\ }\href@noop {} {\bibfield  {journal} {\bibinfo
  {journal} {Physical Review Letters}\ }\textbf {\bibinfo {volume} {87}},\
  \bibinfo {pages} {037901} (\bibinfo {year} {2001})}\BibitemShut {NoStop}%
\bibitem [{\citenamefont {Litinskaya}\ \emph {et~al.}(2016)\citenamefont
  {Litinskaya}, \citenamefont {Tignone},\ and\ \citenamefont
  {Pupillo}}]{litinskaya2016cavity}%
  \BibitemOpen
  \bibfield  {author} {\bibinfo {author} {\bibfnamefont {M.}~\bibnamefont
  {Litinskaya}}, \bibinfo {author} {\bibfnamefont {E.}~\bibnamefont {Tignone}},
  \ and\ \bibinfo {author} {\bibfnamefont {G.}~\bibnamefont {Pupillo}},\
  }\href@noop {} {\bibfield  {journal} {\bibinfo  {journal} {Journal of Physics
  B: Atomic, Molecular and Optical Physics}\ }\textbf {\bibinfo {volume}
  {49}},\ \bibinfo {pages} {164006} (\bibinfo {year} {2016})}\BibitemShut
  {NoStop}%
\bibitem [{\citenamefont {Grankin}\ \emph {et~al.}(2014)\citenamefont
  {Grankin}, \citenamefont {Brion}, \citenamefont {Bimbard}, \citenamefont
  {Boddeda}, \citenamefont {Usmani}, \citenamefont {Ourjoumtsev},\ and\
  \citenamefont {Grangier}}]{Grankin2014}%
  \BibitemOpen
  \bibfield  {author} {\bibinfo {author} {\bibfnamefont {A.}~\bibnamefont
  {Grankin}}, \bibinfo {author} {\bibfnamefont {E.}~\bibnamefont {Brion}},
  \bibinfo {author} {\bibfnamefont {E.}~\bibnamefont {Bimbard}}, \bibinfo
  {author} {\bibfnamefont {R.}~\bibnamefont {Boddeda}}, \bibinfo {author}
  {\bibfnamefont {I.}~\bibnamefont {Usmani}}, \bibinfo {author} {\bibfnamefont
  {A.}~\bibnamefont {Ourjoumtsev}}, \ and\ \bibinfo {author} {\bibfnamefont
  {P.}~\bibnamefont {Grangier}},\ }\href@noop {} {\bibfield  {journal}
  {\bibinfo  {journal} {New Journal of Physics}\ }\textbf {\bibinfo {volume}
  {16}},\ \bibinfo {pages} {043020} (\bibinfo {year} {2014})}\BibitemShut
  {NoStop}%
\bibitem [{\citenamefont {Fleischhauer}\ \emph {et~al.}(2005)\citenamefont
  {Fleischhauer}, \citenamefont {Imamoglu},\ and\ \citenamefont
  {Marangos}}]{fleischhauer2005electromagnetically}%
  \BibitemOpen
  \bibfield  {author} {\bibinfo {author} {\bibfnamefont {M.}~\bibnamefont
  {Fleischhauer}}, \bibinfo {author} {\bibfnamefont {A.}~\bibnamefont
  {Imamoglu}}, \ and\ \bibinfo {author} {\bibfnamefont {J.~P.}\ \bibnamefont
  {Marangos}},\ }\href@noop {} {\bibfield  {journal} {\bibinfo  {journal}
  {Reviews of Modern Physics}\ }\textbf {\bibinfo {volume} {77}},\ \bibinfo
  {pages} {633} (\bibinfo {year} {2005})}\BibitemShut {NoStop}%
\bibitem [{\citenamefont {Cohen-Tannoudji}\ \emph {et~al.}(1998)\citenamefont
  {Cohen-Tannoudji}, \citenamefont {Dupont-Roc},\ and\ \citenamefont
  {Grynberg}}]{cohen1998atom}%
  \BibitemOpen
  \bibfield  {author} {\bibinfo {author} {\bibfnamefont {C.}~\bibnamefont
  {Cohen-Tannoudji}}, \bibinfo {author} {\bibfnamefont {J.}~\bibnamefont
  {Dupont-Roc}}, \ and\ \bibinfo {author} {\bibfnamefont {G.}~\bibnamefont
  {Grynberg}},\ }\href@noop {} {\emph {\bibinfo {title} {Atom-photon
  interactions: basic processes and applications}}}\ (\bibinfo  {publisher}
  {Wiley-VCH},\ \bibinfo {address} {New York},\ \bibinfo {year}
  {1998})\BibitemShut {NoStop}%
\bibitem [{\citenamefont {Pritchard}\ \emph {et~al.}(2010)\citenamefont
  {Pritchard}, \citenamefont {Maxwell}, \citenamefont {Gauguet}, \citenamefont
  {Weatherill}, \citenamefont {Jones},\ and\ \citenamefont
  {Adams}}]{pritchard2010cooperative}%
  \BibitemOpen
  \bibfield  {author} {\bibinfo {author} {\bibfnamefont {J.~D.}\ \bibnamefont
  {Pritchard}}, \bibinfo {author} {\bibfnamefont {D.}~\bibnamefont {Maxwell}},
  \bibinfo {author} {\bibfnamefont {A.}~\bibnamefont {Gauguet}}, \bibinfo
  {author} {\bibfnamefont {K.~J.}\ \bibnamefont {Weatherill}}, \bibinfo
  {author} {\bibfnamefont {M.}~\bibnamefont {Jones}}, \ and\ \bibinfo {author}
  {\bibfnamefont {C.~S.}\ \bibnamefont {Adams}},\ }\href@noop {} {\bibfield
  {journal} {\bibinfo  {journal} {Physical Review Letters}\ }\textbf {\bibinfo
  {volume} {105}},\ \bibinfo {pages} {193603} (\bibinfo {year}
  {2010})}\BibitemShut {NoStop}%
\bibitem [{\citenamefont {Schine}\ \emph {et~al.}(2018)\citenamefont {Schine},
  \citenamefont {Chalupnik}, \citenamefont {Can}, \citenamefont {Gromov},\ and\
  \citenamefont {Simon}}]{Schine2018}%
  \BibitemOpen
  \bibfield  {author} {\bibinfo {author} {\bibfnamefont {N.}~\bibnamefont
  {Schine}}, \bibinfo {author} {\bibfnamefont {M.}~\bibnamefont {Chalupnik}},
  \bibinfo {author} {\bibfnamefont {T.}~\bibnamefont {Can}}, \bibinfo {author}
  {\bibfnamefont {A.}~\bibnamefont {Gromov}}, \ and\ \bibinfo {author}
  {\bibfnamefont {J.}~\bibnamefont {Simon}},\ }\href@noop {} {\bibfield
  {journal} {\bibinfo  {journal} {arXiv:1802.04418}\ } (\bibinfo {year}
  {2018})}\BibitemShut {NoStop}%
\bibitem [{\citenamefont {Simon}(2010)}]{simon2010cavity}%
  \BibitemOpen
  \bibfield  {author} {\bibinfo {author} {\bibfnamefont {J.}~\bibnamefont
  {Simon}},\ }\href@noop {} {\emph {\bibinfo {title} {Cavity QED with atomic
  ensembles}}}\ (\bibinfo  {publisher} {Harvard University},\ \bibinfo {year}
  {2010})\BibitemShut {NoStop}%
\bibitem [{\citenamefont {Ketterle}\ \emph {et~al.}(1999)\citenamefont
  {Ketterle}, \citenamefont {Durfee},\ and\ \citenamefont
  {Stamper-Kurn}}]{ketterle1999making}%
  \BibitemOpen
  \bibfield  {author} {\bibinfo {author} {\bibfnamefont {W.}~\bibnamefont
  {Ketterle}}, \bibinfo {author} {\bibfnamefont {D.~S.}\ \bibnamefont
  {Durfee}}, \ and\ \bibinfo {author} {\bibfnamefont {D.}~\bibnamefont
  {Stamper-Kurn}},\ }\href@noop {} {\bibfield  {journal} {\bibinfo  {journal}
  {arXiv:cond-mat/9904034}\ } (\bibinfo {year} {1999})}\BibitemShut {NoStop}%
\bibitem [{\citenamefont {Goodman}(2005)}]{goodman2005introduction}%
  \BibitemOpen
  \bibfield  {author} {\bibinfo {author} {\bibfnamefont {J.~W.}\ \bibnamefont
  {Goodman}},\ }\href@noop {} {\emph {\bibinfo {title} {Introduction to Fourier
  optics}}}\ (\bibinfo  {publisher} {Roberts and Company Publishers},\ \bibinfo
  {address} {Englewood, CO},\ \bibinfo {year} {2005})\BibitemShut {NoStop}%
\bibitem [{\citenamefont {Grankin}\ \emph {et~al.}(2018)\citenamefont
  {Grankin}, \citenamefont {Grangier},\ and\ \citenamefont
  {Brion}}]{grankin2018diagrammatic}%
  \BibitemOpen
  \bibfield  {author} {\bibinfo {author} {\bibfnamefont {A.}~\bibnamefont
  {Grankin}}, \bibinfo {author} {\bibfnamefont {P.}~\bibnamefont {Grangier}}, \
  and\ \bibinfo {author} {\bibfnamefont {E.}~\bibnamefont {Brion}},\
  }\href@noop {} {\bibfield  {journal} {\bibinfo  {journal} {arXiv:1803.09480}\
  } (\bibinfo {year} {2018})}\BibitemShut {NoStop}%
\bibitem [{\citenamefont {Das}\ \emph {et~al.}(2016)\citenamefont {Das},
  \citenamefont {Grankin}, \citenamefont {Iakoupov}, \citenamefont {Brion},
  \citenamefont {Borregaard}, \citenamefont {Boddeda}, \citenamefont {Usmani},
  \citenamefont {Ourjoumtsev}, \citenamefont {Grangier},\ and\ \citenamefont
  {S{\o}rensen}}]{das2016photonic}%
  \BibitemOpen
  \bibfield  {author} {\bibinfo {author} {\bibfnamefont {S.}~\bibnamefont
  {Das}}, \bibinfo {author} {\bibfnamefont {A.}~\bibnamefont {Grankin}},
  \bibinfo {author} {\bibfnamefont {I.}~\bibnamefont {Iakoupov}}, \bibinfo
  {author} {\bibfnamefont {E.}~\bibnamefont {Brion}}, \bibinfo {author}
  {\bibfnamefont {J.}~\bibnamefont {Borregaard}}, \bibinfo {author}
  {\bibfnamefont {R.}~\bibnamefont {Boddeda}}, \bibinfo {author} {\bibfnamefont
  {I.}~\bibnamefont {Usmani}}, \bibinfo {author} {\bibfnamefont
  {A.}~\bibnamefont {Ourjoumtsev}}, \bibinfo {author} {\bibfnamefont
  {P.}~\bibnamefont {Grangier}}, \ and\ \bibinfo {author} {\bibfnamefont
  {A.~S.}\ \bibnamefont {S{\o}rensen}},\ }\href@noop {} {\bibfield  {journal}
  {\bibinfo  {journal} {Physical Review A}\ }\textbf {\bibinfo {volume} {93}},\
  \bibinfo {pages} {040303} (\bibinfo {year} {2016})}\BibitemShut {NoStop}%
\bibitem [{\citenamefont {Siegman}(1986)}]{Siegman1986}%
  \BibitemOpen
  \bibfield  {author} {\bibinfo {author} {\bibfnamefont {A.~E.}\ \bibnamefont
  {Siegman}},\ }\href@noop {} {\emph {\bibinfo {title} {Lasers}}}\ (\bibinfo
  {publisher} {University Science Books},\ \bibinfo {address} {Mill Valley,
  CA},\ \bibinfo {year} {1986})\BibitemShut {NoStop}%
\end{thebibliography}%

\clearpage
\onecolumngrid
\appendix

\section{Resonator Degeneracies and Photon Localization} \label{SI:photonlocalization}

In Sections ~\ref{atomresonatorcoupling}-~\ref{interactiondrivenloss}, we treat photons as objects that can occupy a completely arbitrary spatial mode in the 2D plane transverse to the resonator axis. In practice, resonator geometry can impose additional symmetries on the allowed photon wave-functions, and limit the permissible degree of photon localization. In what follows we explore the wave-function constraints imposed by various resonator configurations.

\begin{figure}[h]
\centering
	 \subfloat[][]{\includegraphics[scale=0.375,valign=c]{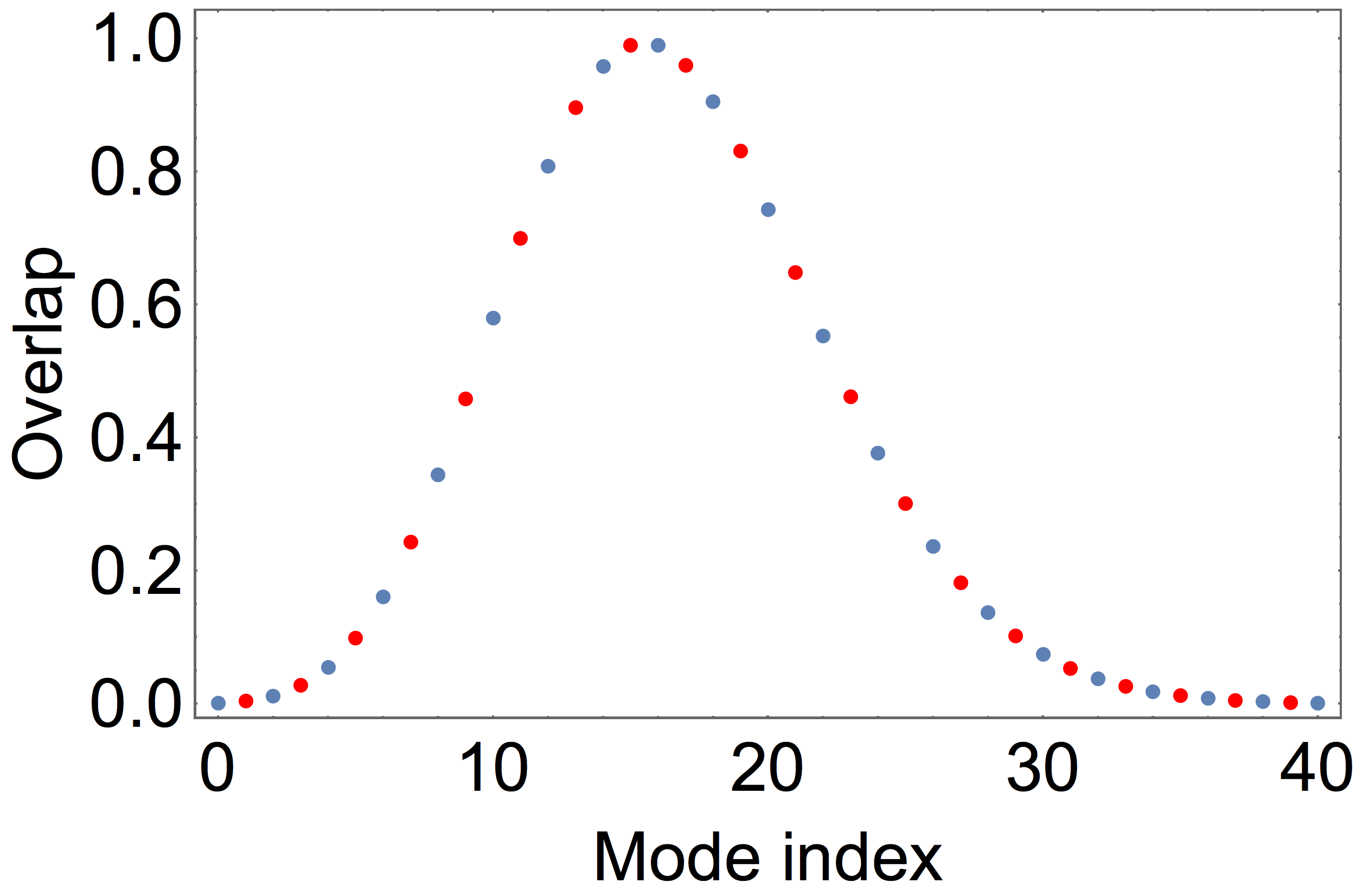}}
     \subfloat[][]{\includegraphics[scale=0.25,valign=c]{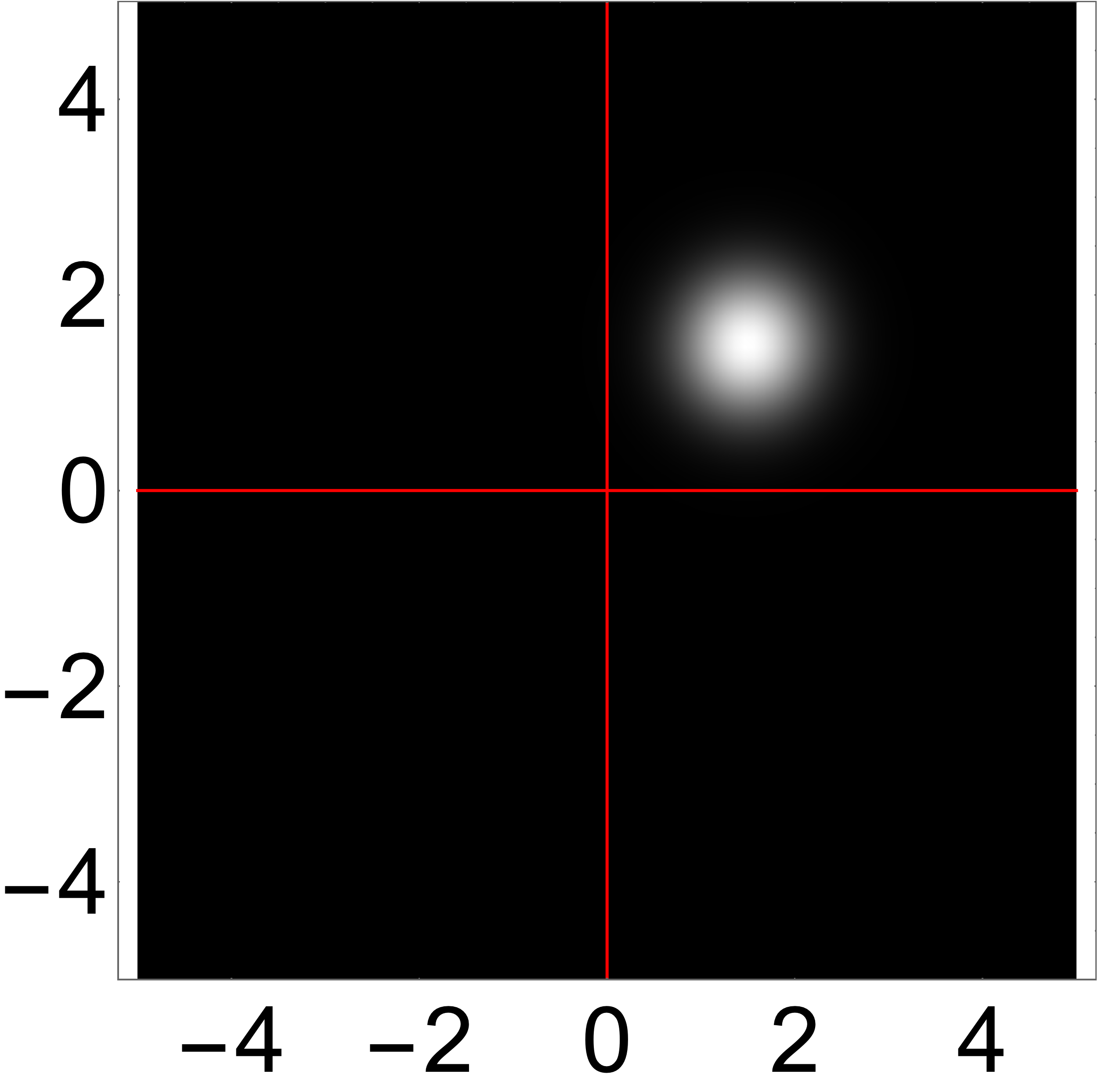}}
     \subfloat[][]{\includegraphics[scale=0.25,valign=c]{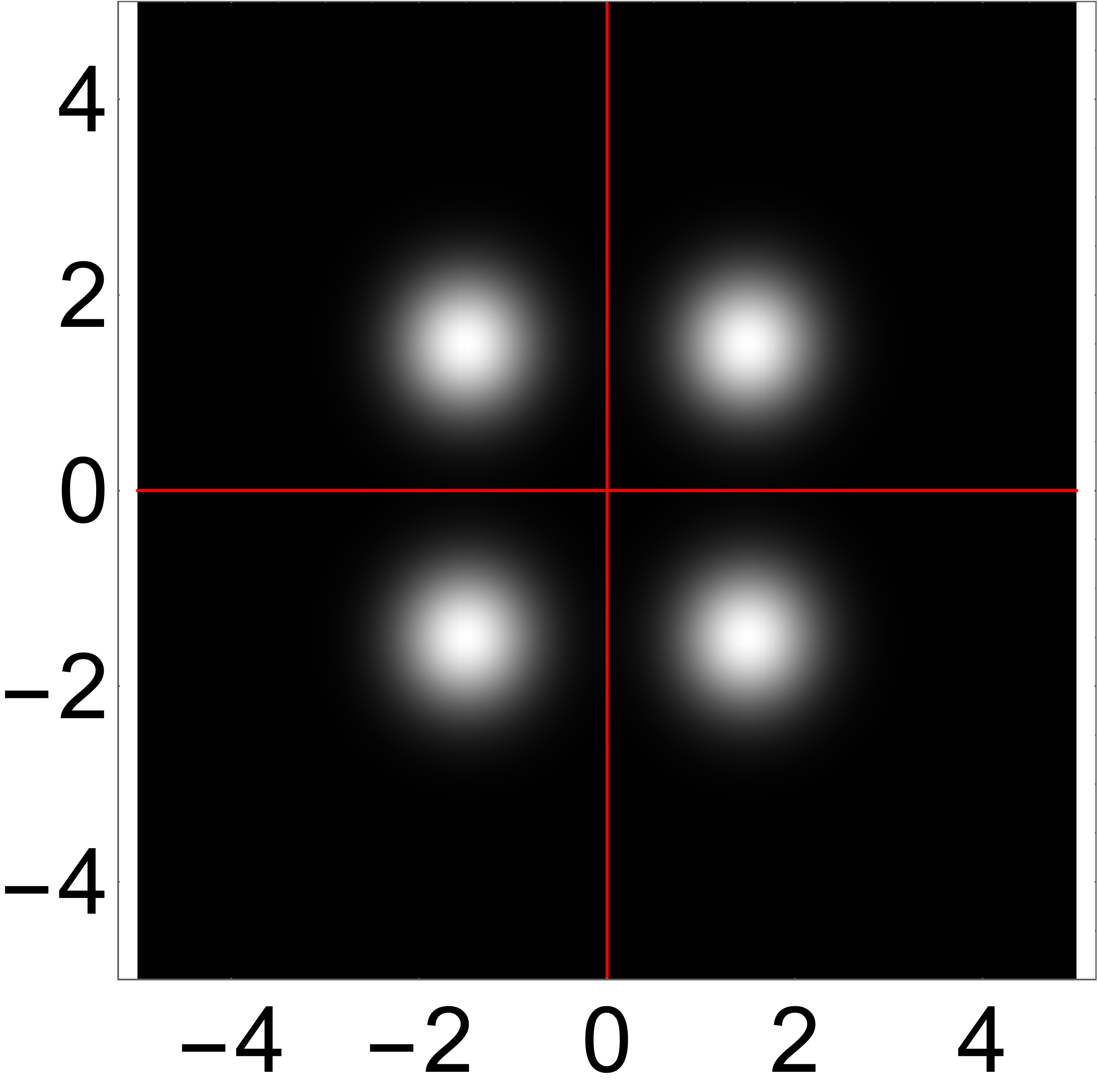}}
     \subfloat[][]{\includegraphics[scale=0.25,valign=c]{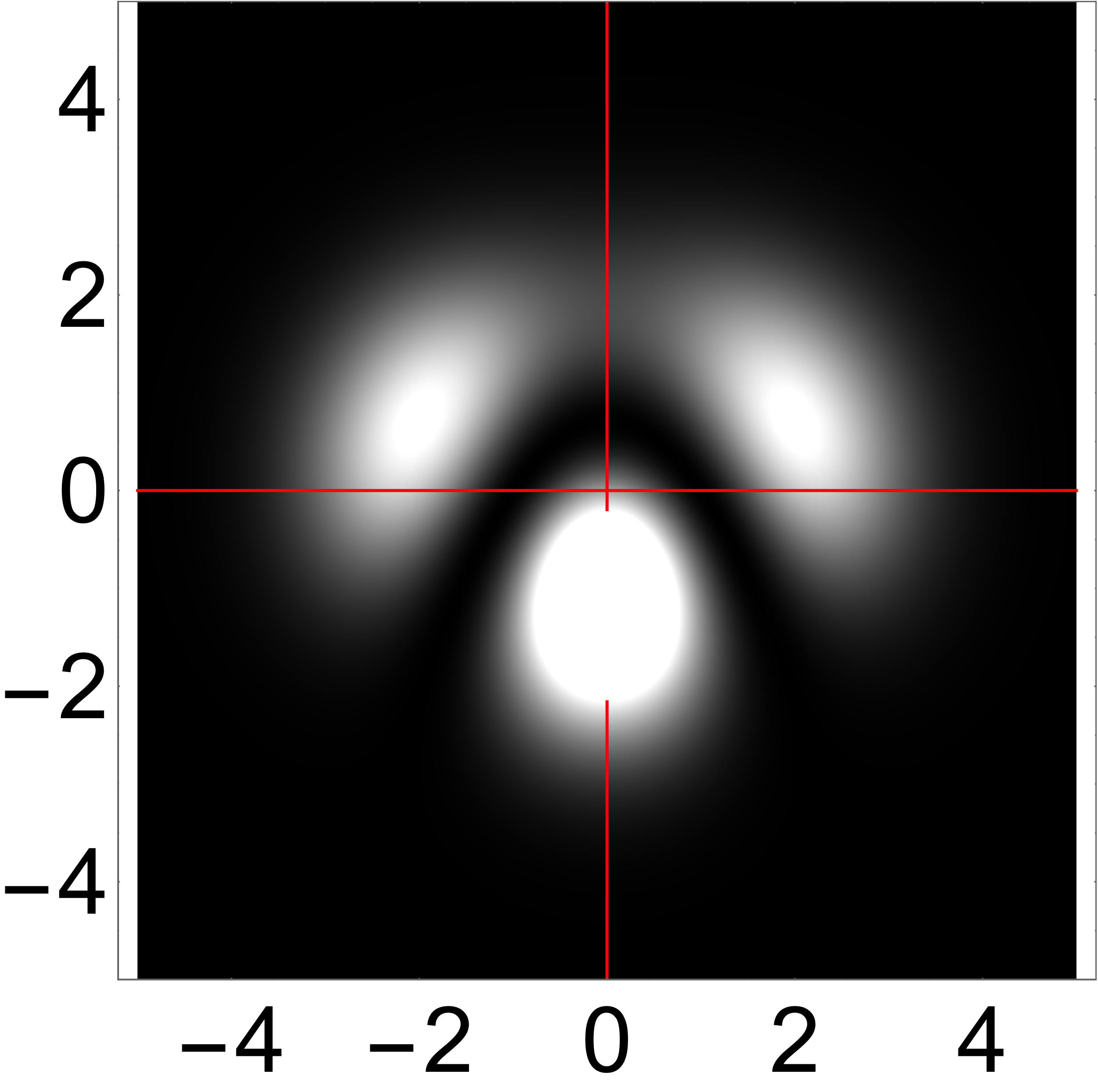}}
     \subfloat[][]{\includegraphics[scale=0.25,valign=c]{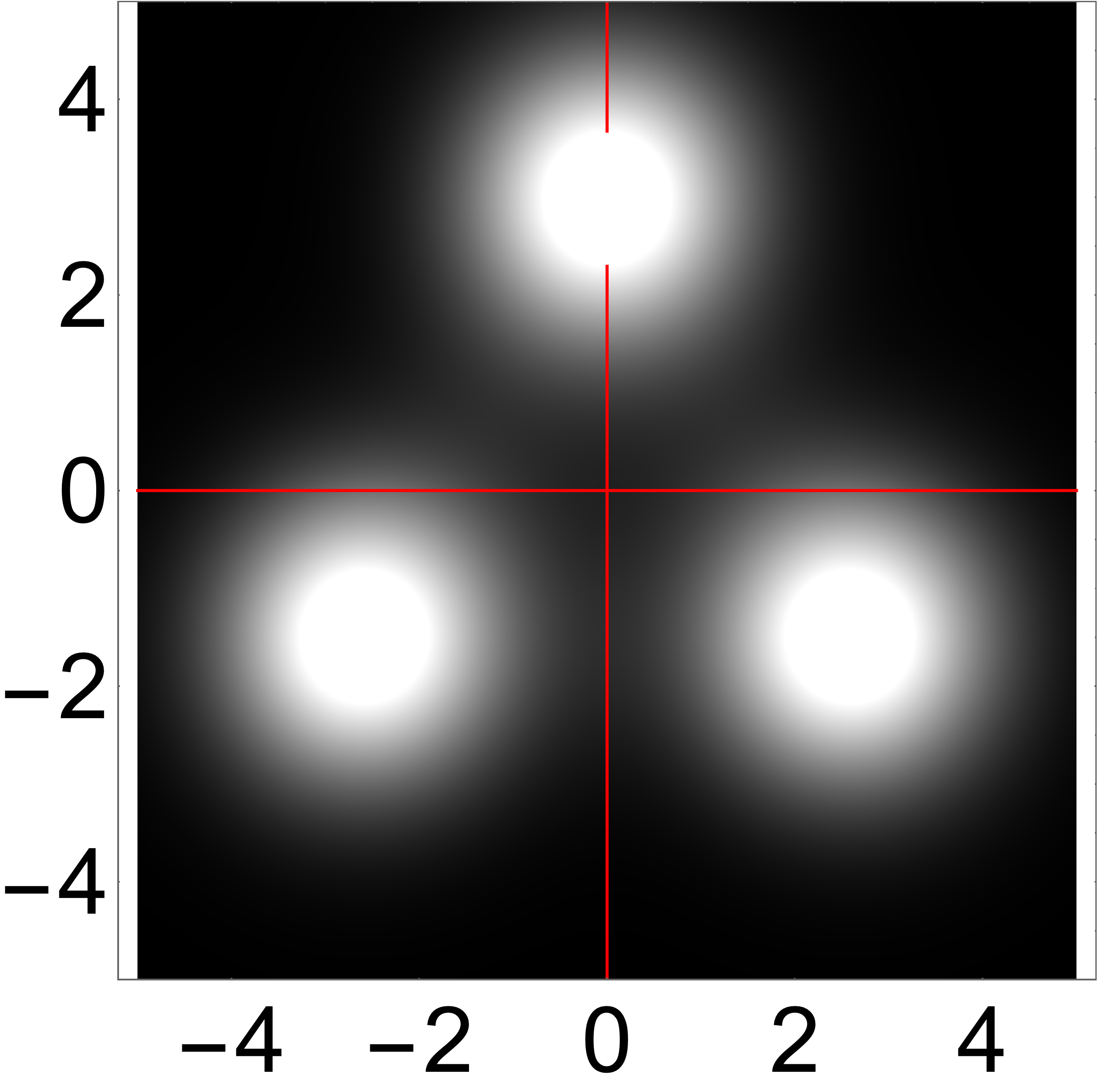}}
     \caption {\textbf{Resonator degeneracies and their effects on photon localization and spot size. (a)}Decomposition of an off-center Gaussian into 1D HG modes. A Gaussian at any arbitrary location can be written as the sum of 1D HG modes, by taking the spatial overlap of the Gaussian with each mode. The overlaps are shown in the plot, with even modes represented as blue dots and odd modes as red dots. \textbf{(b)}Localization of a photon in a planar or concentric cavity. When all resonator modes are degenerate, we can create a photon at any location inside the resonator and of any spot size. This can be understood as a generalized version of the decomposition in (a), but in general it is done in a two dimensional parameter space. The red lines in the plot signify $x=0,y=0$, showing than an off-center spot has been created. \textbf{(c)}In a confocal resonator, we have degeneracy within the even and odd mode manifolds only. As such, a decomposition of any spot would include either the even (blue) or odd modes (red dots) only. Any spot size at any location can still be created but now we pick up reflections through the origin, resulting in a mirrored spot. \textbf{(d)}Superposition of TEM$_{20}$ and TEM$_{01}$. This is an example of a finite one parameter family. The resonator mirrors and mirror positions are picked in such a way as to make these two modes energetically degenerate. The result is interference between the two modes, resulting in a new mode shape~\cite{Jia2016}. \textbf{(e)} Threefold symmetry in a resonator. In a countably infinite single parameter family of degenerate modes where every third LG mode is degenerate (LG$_l$,LG$_{l+3}$,LG$_{l+6}$,...), a localized spot can be created at any point but must satisfy a threefold rotational symmetry, resulting in three copies, each of which cannot be smaller than the LG$_0$ waist size of the resonator~\cite{Schine2016}.}
     \label{resonatordegeneracies}
\end{figure}

It is convenient to begin by considering a general non-degenerate spherical mirror Fabry-perot resonator~\cite{Siegman1986}, whose modes are enumerated with three indices $l,m,n$; the first index $l$, is the longitudinal mode number, while $m,n$ index the transverse mode:

The most extreme case is degeneracy of all the transverse modes of a resonator, $\omega_{lmn}=\omega_l$, achieved in planar and concentric cavities~\cite{Siegman1986} (note that we have re-indexed the longitudinal modes of the concentric resonator). Planar resonators are constructed with flat mirrors (radius of curvature $\rightarrow \infty$) very close together, while concentric cavities consist of two mirrors separated by the sum of their radii of curvature. In both configurations the space of allowed photon wave-functions is not constrained by any symmetries, and arbitrarily small spots can be created at any point in 2D space (see Figure ~\ref{resonatordegeneracies}b).

The next most extreme case is families of degenerate modes that still require two indices to enumerate, but with restrictions on the indices. An example of this is the confocal resonator: two mirrors with radii of curvature $R$ placed a distance R apart. Such a resonator exhibits $\omega_{lmn}=\omega_{l,mod(m+n,2)}$. The constraint that $m+n$ is either even or odd imposes a reflection symmetry across the origin: the photon may be arbitrarily well localized in space at any location, but must \emph{simultaneously} exist at this mirror-image location (see Figure ~\ref{resonatordegeneracies}c).

The next case is degenerate families that may be indexed with only a single parameter, which may themselves be further broken down into two sub-categories: (1) families in which the index takes on only a finite number of values; and (2) families in which the index takes on a countably infinite number of values. A spherical mirror Fabry-Perot falls into the first category; $\omega_{lmn}=\omega_{l,m+n}$; another example is an astigmatic resonator whose length is tuned to enforce degeneracy such as $\omega_{lmn}=\omega_{l,m+2n}$~\cite{Jia2016} (see Figure ~\ref{resonatordegeneracies}d). The latter category could be achieved in an astigmatic resonator tuned to confocality on only one axis: $\omega_{lmn}=\omega_{l,m,mod(n,2)}$, or in a non-planar resonator by imposing twist which is a rational fraction of $2\pi$, as in~\cite{Schine2016}:$\omega_{lmn}=\omega_{l,m,mod(n,3)}$ (see Figure ~\ref{resonatordegeneracies}e). The reduced degeneracy strongly constrains the wavefunctions which may be represented by the family, to the point that the physical interpretation of these families is often quite unclear. Indeed, the twisted resonator explored in ~\cite{Schine2016} exhibits three-fold rotational symmetry, and quantum geometry, meaning that has a minimum spot size; this system may be understood as a Landau level on the surface of a cone, quite an exotic manifold indeed.

\newpage
\clearpage
\newpage

\onecolumngrid

\end{document}